\documentclass[iop]{emulateapj}
\usepackage{apjfonts}
\usepackage{times} 
\usepackage{graphicx}
\usepackage{rotating}
\usepackage{array}
\usepackage[normalem]{ulem}
\usepackage{color}
\usepackage{epsfig}
\usepackage{amsmath}
\usepackage{natbib}
\def\deg{^\circ}
\def\mbh{\ifmmode{{\mathrm M}_{\rm BH}\,}\else{M$_{\rm BH}$\,}\fi}
\def\msigma{\ifmmode{{\mathrm M}_{\rm BH}-{\sigma}\,}\else{M$_{\rm BH}- \sigma$\,}\fi}
\def\msun{\ifmmode{{\mathrm M}_{\odot}}\else{M$_{\odot}$}\fi} 
\def\lsun{\ifmmode{{\mathrm L}_{\odot}}\else{L$_{\odot}$}\fi} 
\def\kms{\ifmmode{{\mathrm{km \, s^{-1}}}}\else{${\mathrm{km \, s^{-1}}}$}\fi}

\def\siglos{\ifmmode{{\mathrm \sigma}_{\rm los}}\else{$\sigma_{\rm los}$}\fi}
\def\vlos{\ifmmode{{\mathrm V}_{\rm los}}\else{$V_{\rm los}$}\fi}
\def\chiall{\ifmmode{{\chi^2}_{\rm all}}\else{$\chi^2_{\rm all}$}\fi}
\def\chikin{\ifmmode{{\chi^2}_{\rm kin}}\else{$\chi^2_{\rm kin}$}\fi}

\def\aj{AJ}
\def\apj{ApJ}
\def\apjl{ApJ}
\def\apjs{ApJS}
\def\aap{A\&A}
\def\mnras{MNRAS}
\def\pasp{PASP}
\def\pasj{PASJ}
\def\procspie{Proc.~SPIE}

\shortauthors{Onken et~al.}
\shorttitle{Black Hole Mass of NGC~4151}

\begin{document}

\title{The Black Hole Mass of NGC~4151. II. Stellar Dynamical Measurement 
from Near-Infrared Integral Field Spectroscopy} 
\author{Christopher A.~Onken\altaffilmark{1,2}, 
Monica Valluri\altaffilmark{3},
Jonathan S. Brown\altaffilmark{3,4},
Peter J.~McGregor\altaffilmark{2},
Bradley M.~Peterson\altaffilmark{4,5},
Misty C.~Bentz\altaffilmark{6},
Laura Ferrarese\altaffilmark{1},
Richard W.~Pogge\altaffilmark{4,5},
Marianne Vestergaard\altaffilmark{7,8},
Thaisa Storchi-Bergmann\altaffilmark{9},
Rogemar A.~Riffel\altaffilmark{10}
} 

\altaffiltext{1}{Herzberg Institute of Astrophysics, National Research
  Council of Canada, 5071 West Saanich Road, Victoria, BC, V9E 2E7,
  Canada}
\altaffiltext{2}{Research School of
  Astronomy \& Astrophysics, The Australian National University,
  Canberra, ACT, 2611, Australia}
\altaffiltext{3}{Department of Astronomy, University of Michigan, 500
  Church St., Ann Arbor, MI 48109-1042, USA}
\altaffiltext{4}{Department of Astronomy, The Ohio State University,
  140 West 18th Avenue, Columbus, OH 43210, USA}
\altaffiltext{5}{Center for Cosmology and AstroParticle Physics, The
  Ohio State University, 191 West Woodruff Avenue, Columbus, OH 43210,
  USA}
\altaffiltext{6}{Department of Physics \& Astronomy, Georgia State University, 
  25 Park Place, Office 610, Atlanta, GA 30303, USA}
\altaffiltext{7}{Dark Cosmology Centre, The Niels Bohr Institute,
  Copenhagen University, Juliane Maries Vej 30, 2100 Copenhagen \O,
  Denmark}
\altaffiltext{8}{Steward Observatory, University of Arizona, 933 North
  Cherry Avenue, Tucson, AZ 85721, USA}
\altaffiltext{9}{Universidade Federal do Rio Grande do Sul, Instituto
  de F\'{i}sica, CP 15051, Porto Alegre 91501-970, RS, Brazil}
\altaffiltext{10}{Universidade Federal de Santa Maria, Departamento de
  F\'{i}sica, Centro de Ci\^{e}ncias Naturais e Exatas, 97105-900,
  Santa Maria, RS, Brazil}

\email{christopher.onken@anu.edu.au,mvalluri@umich.edu}

\begin{abstract}

  We present a revised measurement of the mass of the central
  black hole ($\mbh$) in the Seyfert~1 galaxy NGC~4151. The new
  stellar dynamical mass measurement is derived by applying an
  axisymmetric orbit-superposition code to near-infrared integral
  field data obtained using adaptive optics with the Gemini NIFS
  spectrograph. When our models attempt to fit both the NIFS
  kinematics and additional low spatial resolution kinematics, our results depend sensitively on how $\chi^2$ is computed
  --- probably a consequence of complex bar kinematics that manifest
  immediately outside the nuclear region. The most robust results are
  obtained when only the high spatial resolution kinematic constraints
  in the nuclear region are included in the fit. Our
  best estimates for the BH mass and $H$-band mass-to-light ratio are
  $\mbh \sim 3.76\pm1.15 \times 10^7~\msun$ (1$\sigma$ error) and
  $\Upsilon_H \sim 0.34 {\pm 0.03}~\msun/\lsun$ (3$\sigma$ error),
  respectively (the quoted errors reflect the model
  uncertainties). Our BH mass measurement is consistent
  with estimates from both reverberation mapping
  ($3.57^{+0.45}_{-0.37}\times~10^7~\msun$) and gas kinematics
  ($3.0^{+0.75}_{-2.2}\times~10^7~\msun$; 1$\sigma$ errors), and our
  best-fit mass-to-light ratio is consistent with the photometric
  estimate of $\Upsilon_H = 0.4\pm0.2~\msun/\lsun$. The NIFS
  kinematics give a central bulge velocity dispersion $\sigma_c=116\pm
  3$~\kms, bringing this object slightly closer to the \msigma
  relation for quiescent galaxies. Although NGC~4151 is one of only a
  few Seyfert~1 galaxies in which it is possible to obtain a direct
  dynamical BH mass measurement --- and thus, an independent
  calibration of the reverberation mapping mass scale --- the complex
  bar kinematics makes it less than ideally suited for this purpose.
\end{abstract}



\section{INTRODUCTION}\label{sec:intro}

Long before the development of General Relativity, John Michell
wondered about the gravitational influence of objects on the light
they emit, and how one might go about finding objects so dense that
light could not escape their surfaces. ``[I]f any other luminous
bodies should happen to revolve about them we might still perhaps from
the motions of these revolving bodies infer the existence of the
central ones with some degree of probability, as this might afford a
clue to some of the apparent irregularities of the revolving bodies,
which would not be easily explicable on any other hypothesis''
\citep{Michell1784}. More than 200 years later, precisely this method
has been employed to determine the mass of the black hole (BH) at the
center of our Milky Way. Sgr~A$^{\ast}$ is one of the most
tightly constrained BHs in the universe, with a total mass uncertainty
(statistical and systematic) of less than 10\% \cite[see][and
references therein]{Gillessen09}.

Using stellar motions to measure the BH masses in more distant
galaxies is complicated by our current inability to spatially resolve
the orbits of individual stars. Instead, one must rely on the
luminosity-weighted line-of-sight velocity distribution (LOSVD) within
different spatial resolution elements, and model the most likely
contribution to the gravitational potential from the unseen BH (beyond
that provided by the stars themselves and any dark matter). The most
common numerical approach to constraining the BH mass ($\mbh$) and
mass-to-light ratio of the stars ($\Upsilon$) is the orbit
superposition method of \citet{schwarzschild_79}, which simultaneously
optimizes the fit to the surface brightness distribution and the
line-of-sight kinematics of stars within the nuclear
region. Application of that technique has yielded more than 50 BH mass
estimates \citep[e.g.,][]{graham_etal_11,woo_etal_13}.

A complimentary mass scale has been developed for the BHs in active
galactic nuclei (AGNs), based on the technique of reverberation
mapping \citep[RM;][]{Blandford82,Peterson01}. Measuring the
size-scale and velocity of gas near the BH has provided mass estimates
for $\sim$50 BHs \citep{Peterson04,Bentz10}, but systematic
uncertainties remain due to the unknown geometry and dynamics of the
gas. This uncertainty is traditionally encapsulated in a scaling (or
projection) factor, $f.$ An empirical calibration for the RM mass
scale $f$ has been established by assuming that those AGNs also lie on
the tight correlation between BH mass and bulge stellar velocity
dispersion (the $M_{\rm BH}-\sigma$ relation) found in quiescent
galaxies. This was first done by \citet{onken_etal_04}, who obtained
an average RM mass scale of $\langle{f}\rangle=5.5$. Subsequent
determinations have ranged from
$\langle{f}\rangle=5.24^{+1.36}_{-1.07}$ \citep{Woo10} to
$\langle{f}\rangle=2.8^{+0.7}_{-0.5}$ \citep{graham_etal_11}, with the
most recent value being $\langle{f}\rangle=4.31\pm1.05$
\citep{grier_etal_13}.  For NGC~4151, this gives a current
RM-derived BH mass of 
$\mbh=3.57^{+0.45}_{-0.37}\times~10^7~\msun$\ \citep[based on the
revised estimate of][]{grier_etal_13}. It is important to
note that the error generally quoted on a RM-derived BH mass in an
individual galaxy is the formal uncertainty in the virial product and
does not include the uncertainty in the nominal mean scale factor
$\langle{f}\rangle$.

What has been missing is an independent BH mass measurement for a
sample of galaxies with RM estimates. Early results have suggested
that the empirically calibrated RM masses are consistent with the
values produced by other techniques
\citep{davies_etal_06,onken_etal_07}, but the few AGNs near enough to
be probed by stellar dynamics have not yielded particularly tight
constraints (in part, owing to the complicating factor of the bright
AGN overwhelming the stellar absorption features at the smallest
galactic radii). While only a few galaxies lend themselves to
  multiple methods for measuring the masses of their central BHs, such
  independent BH mass determinations are crucial to isolating the
  systematic errors inherent in each method. It is only via
  independent measurements that it is possible to derive robust BH
  masses.

NGC~4151 is a particularly interesting galaxy since it is one of
  the rare objects for which there already exist two independent
  dynamical measurements for the mass of the BH. \citet{onken_etal_07}
  use optical spectroscopy of stellar absorption lines and orbit
  superposition modeling to derive an upper limit for the BH mass of
  $4\times 10^7\msun$ when the galactic bulge is assumed to be
  edge-on. When the bulge is assumed to have the same inclination as
  the large-scale disk, they derive a BH mass of 4-5$\times
  10^7\msun$, which they caution is likely to be a biased solution
  since it is associated with a very noisy $\chi^2$ surface that also
  gives a poor fit to the data (in terms of absolute $\chi^2$ value).

\citet{hicks_malkan_08} use adaptive optics (AO) to obtain high
  spatial resolution, near-infrared spectroscopy of the molecular,
  ionized, and highly ionized gas in NGC~4151. From the kinematics of
  gas in the vicinity of the BH, they derive a dynamical mass of
  $\mbh = 3^{+0.75}_{-2.2}\times 10^7\msun$ for the BH. Being somewhat
  smaller than the value obtained by \citet{onken_etal_07}, this
  suggests a need to revisit the stellar dynamical BH mass
  measurement.

Two serious concerns with the result of \citet{onken_etal_07}
  motivate the need for the high spatial resolution observations that
  are presented in this paper. First, based on the RM mass and the BH
  mass derived by \citet{hicks_malkan_08}, the sphere of influence of
  the BH is estimated to be $\approx 0\farcs3$, significantly smaller
  than the spatial resolution of the spectra used by
  \citet{onken_etal_07}. Second, low spatial resolution integral field
  spectroscopy of a larger field of view \citep{dumas_etal_07} shows
  that the true kinematic major axis of the bulge (derived from the 2D
  velocity field) is oriented $\sim 70\deg$ from the axis assumed by
  \citet{onken_etal_07}.  The observations and modeling presented in
  this paper are designed to overcome both of these drawbacks and to
  provide a more robust comparison with the gas dynamical measurement
  of \citet{hicks_malkan_08}.

To enhance the comparison between the RM and stellar dynamical mass
scales for BHs, we have obtained near-IR integral field spectroscopy
with AO in order to study the stellar dynamics in the
inner regions of the reverberation-mapped AGN, NGC~4151. In
conjunction with our previous optical spectroscopy and
multi-wavelength imaging data, we are able to place the tightest
dynamical constraints to date on the BH mass in NGC~4151. In
Section~2, we describe the observational data. In Section~3, we
discuss our stellar dynamical modeling. Section~4 describes our
results. Section~5 contains further discussion and our conclusions.

\section{OBSERVATIONS}
\label{sec:obs}

NGC~4151 is a nearby ($z=0.003319$) galaxy with a prominent bulge,
faint spiral arms and a large scale bar (see Sections 3.4 and 3.5 for
more details on the morphology and distance of the galaxy). The
galaxy's nuclear activity has been studied extensively since the
original paper by Seyfert (1943); in particular, the analysis of UV
and optical emission lines have been used for determining the BH mass
via RM \citep[e.g.,][]{bentz_etal_06}.
The analysis presented here uses new near-IR spectroscopy performed
with AO, as well as previously published optical long-slit
spectroscopy, and space- and ground-based imaging \cite[primarily
from][hereafter Paper~I]{onken_etal_07} to provide a mass estimate
independent of the RM value.

\subsection{Near-IR Spectroscopy}
\label{sec:nifs_details}

Observations of NGC~4151 and three giant stars (used as velocity
templates) were obtained with the Near-infrared Integral Field
Spectrograph \cite[NIFS;][]{mcgregor_etal_03} on Gemini North between
16 Feb 2008 and 24 Feb 2008 in queue mode (Program ID GN-2008A-Q-41).
The NIFS image-slicing integral field unit (IFU) provides $R\sim5000$
near-IR spectra across a $2\farcs99\times2\farcs97$ field, with a
spatial sampling of $0\farcs103\times0\farcs043$, or $29\times69$
``spaxels'' (spatial pixels), where the smaller pixel scale is along
each of the 29 image slices. Because previous observations of the
K-band stellar absorption lines in NGC~4151 show the features to be
weak in the central regions \citep{ivanov_etal_00,hicks_malkan_08}, we
focus on the CO~(6-3) bandhead at a rest-frame wavelength of
1.62~$\mu$m, which has been found to be quite prominent in previous
NIFS spectra \citep{Storchi-Bergmann09,Riffel09}. The $H$-band spectra
cover a wavelength range of $\sim1.49~\mu$m$-1.80~\mu$m (centered at
1.6490~$\mu$m), with a dispersion of 1.60~\AA~pixel$^{-1}$ across the
2040 spectral pixels, and a velocity resolution of 56.8~\kms.

To take advantage of the spatial sampling afforded by NIFS, the
observations were taken with the Altair AO system \citep{herriot98},
using the AGN in NGC~4151 as the AO guide star. The observations
employed the Altair field lens \citep{boccas06} to provide a more
uniform point spread function (PSF) across the NIFS field. The level
of AO correction depends upon the prevailing seeing conditions at the
time. The median uncorrected seeing was $\approx0\farcs5$, and more
than 95\% of the data were obtained in seeing better than $0\farcs8$,
which provided an AO-corrected nuclear FWHM between $0\farcs10$ and
$0\farcs40$. To estimate the PSF, we model the contribution from
  the AGN to the spectrum in each spaxel across the field of view,
  taking advantage of the AGN's strong [\ion{Fe}{2}]~$1.644~\mu$m
  emission line, at an observed wavelength of
  $\sim1.649\mu$m. Although the [\ion{Fe}{2}] emission is spatially
  extended in NGC~4151, the off-nuclear emission line is narrower than
  the nuclear line profile, and tight constraints can be placed on the
  amount of AGN spectrum that ends up in each spaxel \cite[see \S~4.2
  of][]{jahnke_etal_04}. The analysis was restricted to a wavelength
  region around the [\ion{Fe}{2}] line ($1.6359~\mu$m$-1.6679~\mu$m),
  and to control the effects of noise, the spectra were box-car
  smoothed over 5 pixels. A slice was taken through the center of the
  resulting 2D map, cutting along the smaller pixel scale
  ($0\farcs043$). The PSF profile shows the standard core+halo
  structure delivered by AO systems, which we model as a 2-part
  Gaussian, where the core has $\sigma=0\farcs04$, the halo has
  $\sigma=0\farcs2$, and each component has equal amplitude.

The data for NGC~4151 were acquired at a position angle of $-15\deg$,
which aligns the NIFS slitlets perpendicular to the radio jet
\cite[matching the orientation used
by][]{Storchi-Bergmann09,Riffel09,storchi-bergmann10}. The typical
observing sequence included a 9-point dither pattern with separations
of 0\farcs206, and offsets to a sky field (210\arcsec\ away)
approximately every third frame. Individual exposures were 120~s long,
and the observations were split into sequences of 80 minutes in
duration (in some instances, shortened by changing weather
conditions). Arc lamps were taken in the middle of each observing
sequence, and \ion{A0}{5} telluric standards were obtained before and
after each sequence. The telluric stars were chosen to be 2MASS
sources to simultaneously provide photometric calibration: HD~98152
was observed before NGC~4151, and HD~116405 afterwards. Darks, flats,
and Ronchi mask calibrations were obtained during the daytime after
each night's science observations. In total, 253 on-source frames were
obtained, for a total exposure time of 30360~s. A total of 112 sky
frames were also acquired.

The giant stars we observed were included so as to provide a measure
of the spectral broadening of the instrument, allowing the
intrinsically narrow-featured stars to serve as templates that could
be broadened to match the actual stellar velocity distribution of
NGC~4151. These velocity template stars were observed without AO, and
were selected to span a range of spectral type: HD~35833 (G0),
HD~40280 (\ion{K0}{3}), HIP~60145 (M0). The observations of each star
consist of four on-source and two sky exposures, each with an
integration time of 5.3~s.

\subsubsection{Data Reduction}

The NIFS data were reduced with a combination of version 1.9.1 of the
Gemini IRAF
package\footnote{\url{http://www.gemini.edu/sciops/data-and-results/processing-software}}
and our own IRAF scripts. Calibrations were processed independently
for each night. The observations of NGC~4151 were grouped by proximity
to a telluric standard, typically dividing the observing sequence in
half. The data were flat-fielded, dark-subtracted,
wavelength-calibrated, and spatially rectified, then combined,
corrected for telluric features, and flux-calibrated. The flux
calibration provided by the two telluric/flux standards differs
systematically by $\approx30$\%, implying a 0.3~mag discrepancy in the
2MASS photometry for the two stars. Because the absolute flux levels
in the NIFS data play no role in our subsequent analysis, we do not
attempt to correct for this offset.

Since the image quality of some frames is significantly worse than
others, we select a subsample of individual frames with FWHM values
below a certain upper limit. With a threshold of 0\farcs16 in the FWHM
(measured at a uniform wavelength of 1.6598~$\mu$m to avoid emission
features in the spectra), 236 frames are included in the final
datacube. We spatially shift the data to account for the dither
pattern and then median-combine the images, yielding a datacube
containing 29$\times$69~spatial pixels and 2040 spectral pixels.

The near-IR spectra at each spatial position in the final NIFS
  datacube are combined onto a bin of size $\sim$0\farcs2-square
  (2$\times$5 raw spaxels), resulting in a cube of 15$\times$15
  spaxels. This choice of binning the spaxels arose from two main
  considerations. First, prior to carrying out the axisymmetric
  dynamical modeling, the kinematic data need to be symmetrized about
  the kinematic major axis and reflected about the rotation axis, to
  avoid modeling inconsistencies. This process is best accomplished
  with square bins, especially when the symmetry axes of the galaxy
  are not aligned with the axes of the spectrograph (as is generally
  the case). Second, since the halo of the PSF has $\sigma=0\farcs2$
  and this component comprises 50\% of the overall PSF amplitude, the
  spatial resolution of the data is not significantly altered by using
  larger bins.

We reduce the observations of the telluric/flux standard stars and the
velocity template stars with procedures similar to those we apply to
the NGC~4151 data. We median-combine the four frames for each star,
extract spectra with a fixed aperture of 1$\farcs$5, and then we
normalize each extracted spectrum for the velocity template stars by a
low-order fit to the continuum.

\subsection{Optical Spectroscopy}

In addition to the new data from NIFS, we utilize the results from the
optical spectra of NGC~4151 that were analyzed in Paper~I. The data
consist of long-slit spectra of the Calcium triplet region
($\sim8500$\AA) taken with the Mayall 4~m telescope at Kitt Peak
National Observatory in 2001, and with the 6.5~m MMT Observatory at
Mt.\ Hopkins in 2004. For the data from Kitt Peak, the
  spectroscopic slit was 2\arcsec wide and was placed at a position
  angle of 135$\deg$. To enhance the signal-to-noise, spectra were
  extracted with an aperture that increased in size away from the
  galaxy center, extending to radii of $\pm$14\arcsec. The spectral
  resolution was $\simeq60$~\kms. For the observations with the MMT,
  the slit was 1\arcsec\ wide, at a position angle of
  69$\deg$. Spectra were extracted in $0\farcs3$ bins to a distance of
  12\arcsec on either side of the AGN. The spectral resolution was
  $\simeq50$~\kms. The seeing conditions for the two datasets were
  $\sim$1.8\arcsec\ and $\sim$3\arcsec, respectively. We refer the
reader to Paper~I for full details.

\subsection{Near-IR Imaging}

To model the surface brightness profile of NGC~4151 in the near-IR, we
make use of the $H$-band imaging from the Early Data
Release\footnote{\url{http://www.astronomy.ohio-state.edu/~survey/}}
of the Ohio State University Bright Spiral Galaxy Survey
\cite[OSUBSGS;][]{eskridge_etal_02}. The data were obtained with the
1.8~m Perkins telescope at Lowell Observatory. The image has a plate
scale of 1\farcs5~pixel$^{-1}$, and we estimate the seeing to be
2\arcsec. We photometrically calibrate the OSUBSGS imaging using 2MASS
point-sources in the field, deriving a photometric zero-point of
$H=22.2\pm0.1$~mag.

\subsection{Optical Imaging}

To bridge the gap in the surface brightness information between the
small field-of-view of the AO-assisted NIFS data and the large-scale
low resolution imaging data of the OSUBSGS, we use the {\it Hubble
  Space Telescope} images from Paper~I. These data were taken with the
F550M filter on the High Resolution Channel of the Advanced Camera for
Surveys (ACS), and cover a field-of-view of
$\sim$25\arcsec$\times$29\arcsec. We refer the reader to Paper~I for
additional details.

We also make use of $g$- and $i$-band optical imaging from the Sloan
Digital Sky Survey (SDSS\footnote{\url{http://www.sdss.org/}}) to
assist in constraining the mass-to-light ratio ($\Upsilon$). The
images have an exposure time of 53.9~s in each band, with 0\farcs396
pixels and $\sim$1\farcs0-seeing in both filters. We use the
photometrically calibrated and sky-subtracted images that SDSS makes
available.

\section{Analysis}
\label{sec:analysis}

As in Paper~I, our modeling relies on simultaneously fitting both the
luminosity distribution of the bulge and the observed integrated
line-of-sight velocity distributions (LOSVD) of stars obtained with
various spectrographs over a range of spatial scales. We detail below
the methods of deriving those quantities from the data described
above.

\subsection{Stellar Velocity Field}

As discussed in \ref{sec:nifs_details}, the NIFS instrument
  produces a grid of spatial pixels (``spaxels'') of size
  $0\farcs103\times0\farcs043$ which were combined onto a bin of size
  0\farcs2-square, resulting in a cube of 15$\times$15 spaxels. The
spectrum from each binned spaxel is then analyzed using the
``penalized pixel fitting'' (pPXF) package\footnote{Available at
  \url{http://purl.org/cappellari}} for
IDL\footnote{\url{http://www.exelisvis.com/ProductsServices/IDL.aspx}}
by \citet{cappellari_emsellem_04}. The pPXF routine computes the
velocity shift, the velocity dispersion, and the higher-order moments
of the stellar LOSVDs (parameterized as Gauss-Hermite (GH)
polynomials) that, when convolved with the velocity template star
spectrum, provide the best match to the observed data. The software
allows emission features to be masked in the fit, and also provides
for multiplicative and additive polynomials to be included (which we
use to account for the contribution of the AGN continuum).

\begin{figure*}
\centering \includegraphics[trim=0.pt 0.pt 0.pt 0.pt, clip, width=1.0\textwidth]{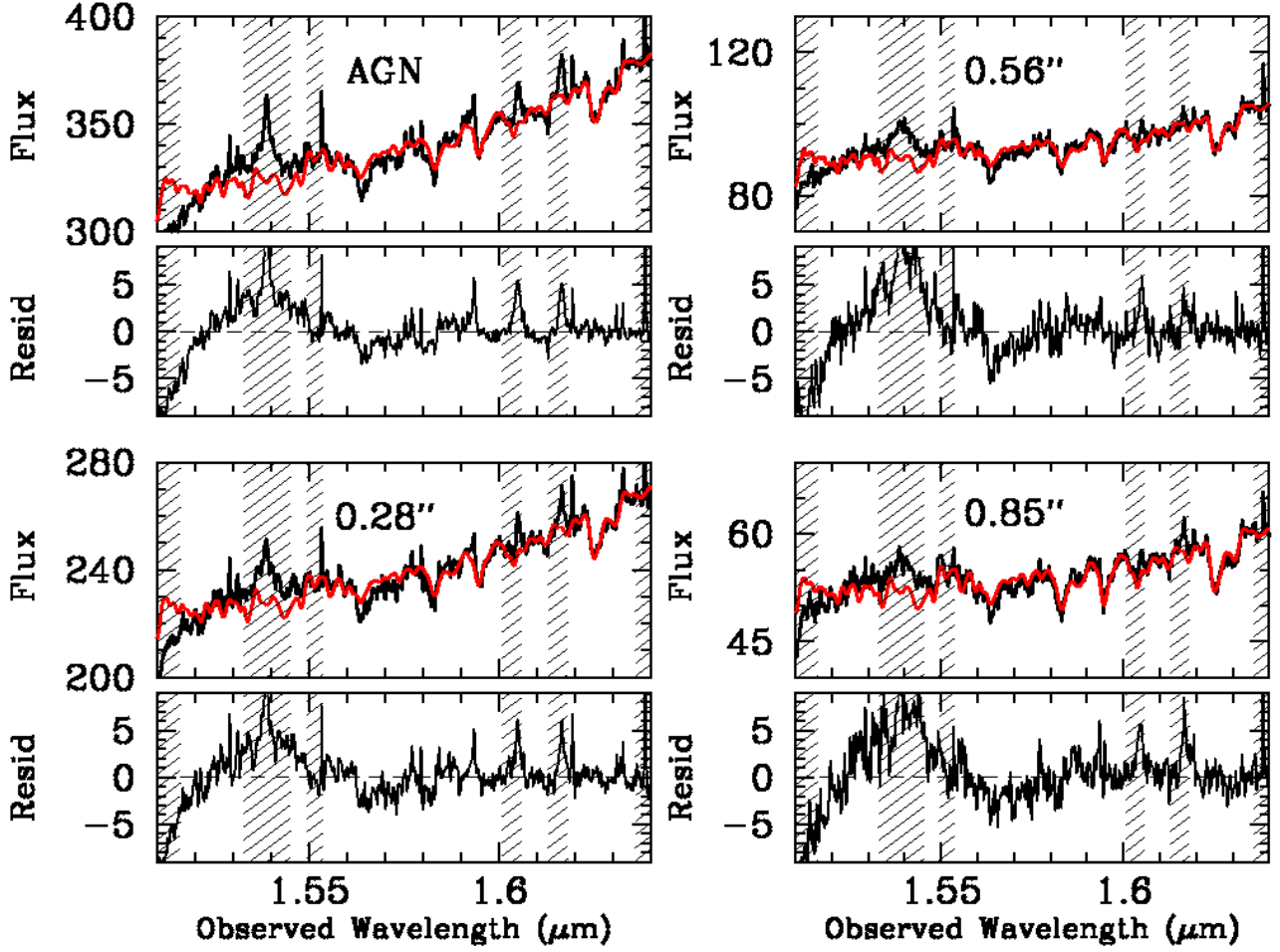}
\caption{Spectra from four spaxels of the rebinned NIFS datacube, with
  fluxes in arbitrary units (although the relative flux scaling
  between spaxels is correct). The best-fit spectra, obtained by
  convolving the LOSVDs from pPXF with the M0 template spectrum, are
  overplotted in light grey (colored red in the online version). The
  lower panel in each pair shows the residuals from each fit, in
  percentage units. The cross-hatched wavelength regions were excluded
  from the pPXF fitting in order to avoid contaminating spectral
  features (primarily AGN emission lines). The four pairs of panels
  are: ({\it upper left}) the central spaxel, where the AGN provides
  the majority of the flux; ({\it lower left}) a spaxel 0$\farcs$28
  from the BH; ({\it upper right}) a spaxel 0$\farcs$56 from the BH;
  ({\it lower right}) a spaxel 0$\farcs$85 from the BH. }
\label{fig.spec}
\vspace{0.2cm}
\end{figure*}

Our pPXF analysis fits the LOSVD up to 4th-order GH terms, including
2nd-order multiplicative and additive polynomials, and fits the
spectra over the wavelength range $1.51625~\mu$m$-1.63562~\mu$m, with
four small wavelength gaps around AGN emission lines and residual sky
features. Figure~\ref{fig.spec} shows four example spectra and pPXF
fits, from different positions within the datacube. We obtain the
smallest uncertainties on the kinematic parameters when we fit
the AGN spectra with the M0 velocity template, and allowing
simultaneous use of all three templates does not measurably improve
the results. We also compared the velocity measurements to those
obtained when using NIFS velocity templates (spectral types
\ion{K0}{3}, \ion{K5}{3}, \ion{M1}{3}, and \ion{M5}{1}a) acquired by
Watson et al.\ (2008; Gemini program GN-2006B-SV-110). The M-type
  stars provided better matches to the relative strengths of the
  absorption lines in each NGC~4151 spectrum, and the kinematic
  uncertainties were smaller than for the K-type stars. Compared to
  the kinematics derived from our M0 velocity template, the
  differences were typically less than 1$\sigma$, suggesting our fits
  are robust against the effects of template-mismatch\footnote{While
    the optical kinematics from Paper~I were derived using a
    \ion{K3}{3} star, the Calcium triplet has been shown to be
    insensitive to template-mismatch \citep{barth02}.}. Our results
  are consistent with those of \citet{kang_etal_13}, who found better
  matches to $H$-band galaxy spectra with M-type stars than with
  K-type stars.

\begin{figure*}
\centering \includegraphics[trim=0.pt 130.pt 0.pt 30.pt, clip, width=1.\textwidth]{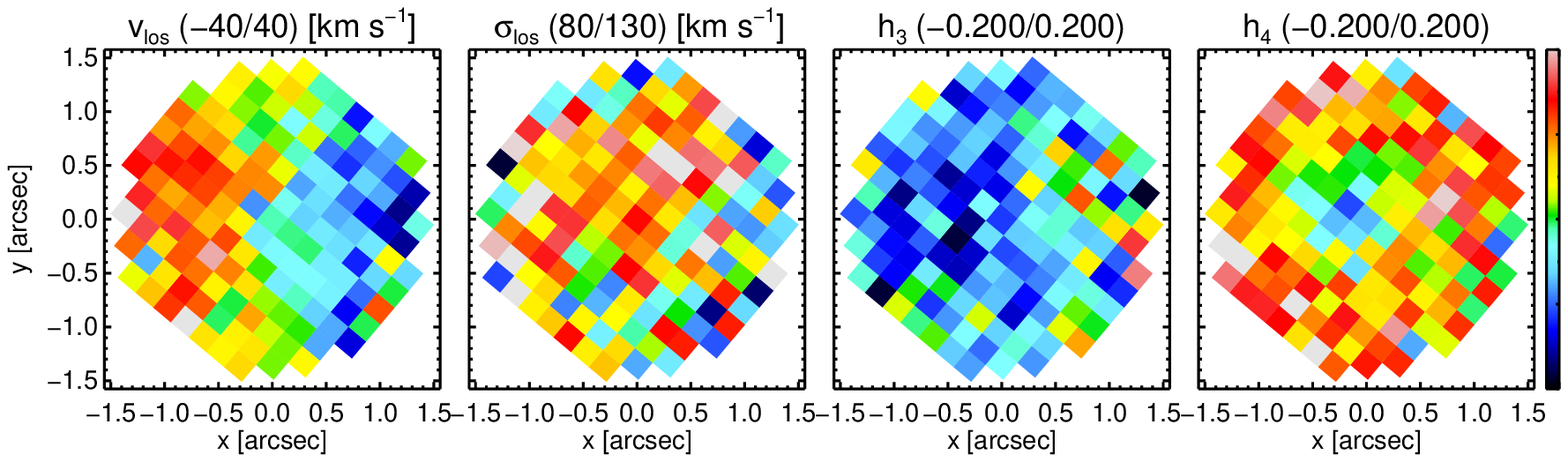}
\centering \includegraphics[trim=0.pt 130.pt 0.pt 30.pt, clip, width=1.\textwidth]{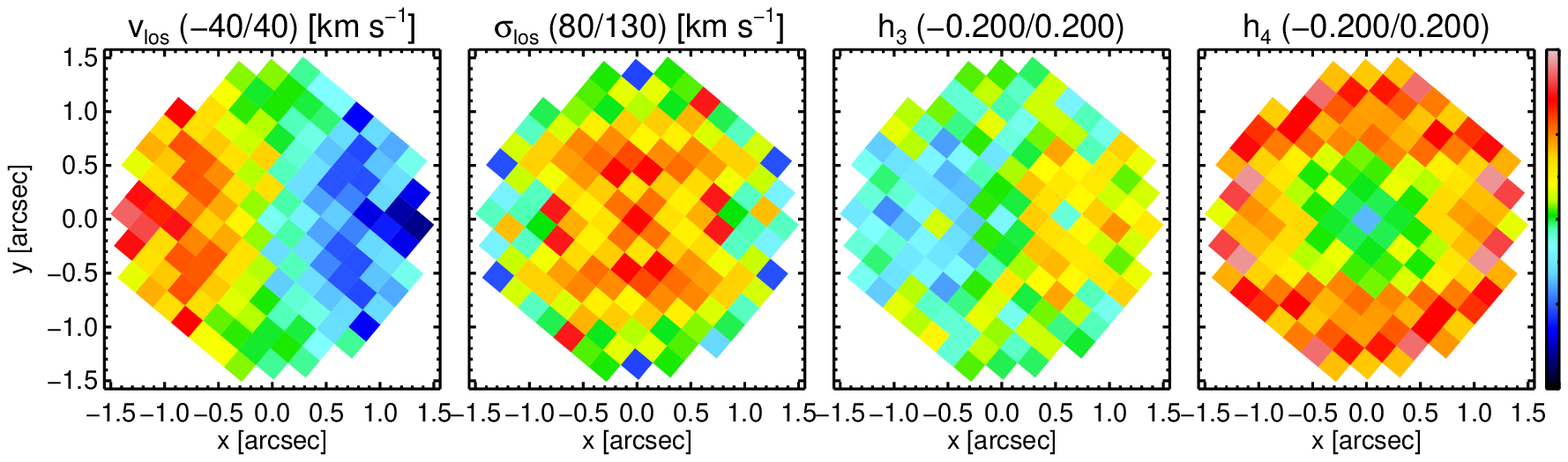}
\caption{Maps of line-of-sight velocity (\vlos), velocity dispersion
  (\siglos) and the 3rd and 4th Gauss-Hermite (GH) moments ($h_3,
  h_4$) in 0\farcs2 spaxels. The $x,y$ coordinates are aligned with
  the major and minor kinematic axes, respectively. The minimum and
  the maximum values of the color scale are indicated in the title of
  each panel in parenthesis ($V_{\rm los}, \sigma_{\rm los}$ in km/s;
  $h_3, h_4$ in dimensionless units). The top row shows the original
  kinematical data while the bottom row shows the kinematics after it
  is bi-symmetrized. }
\label{fig:vel_fields_orig}
\vspace{0.2cm}
\end{figure*}

Figure~\ref{fig:vel_fields_orig}~(top row) shows maps of the
line-of-sight velocity ($V_{\rm los}$), velocity dispersion
($\sigma_{\rm los}$), and the 3rd and 4th GH moments ($h_3, h_4$) in
0\farcs2 spaxels. The $x,y$ coordinates are aligned with the major and
minor kinematic axes, respectively. The color scale is in km/s for
$V_{\rm los}$ and $\sigma_{\rm los}$, and in dimensionless units for
$h_3$ and $h_4$.

We then bi-symmetrize the 2D LOSVD map
(Fig.~\ref{fig:vel_fields_orig}~bottom row) around the kinematic minor
axis at a position angle of $-65\deg$ ($-50\deg$ relative to the NIFS
y-axis), using an IDL routine developed by
\citet{vandenbosch_dezeeuw_10}.  This procedure enhances the S/N by
combining the measurements on either side of the kinematic minor axis
in a symmetric ($\sigma_{\rm los}$ and $h_4$), or anti-symmetric
($V_{\rm los}$ and $h_3$) way.\footnote{For example, an $h_3$ data
  point is multiplied by -1 and averaged with the $h_3$ value at a
  position mirrored on the opposite side of the BH.} Such a procedure
is necessary in order to obtain sensible results from an axisymmetric
dynamical modeling code, which assumes such an underlying
symmetry. The kinematic data were bi-symmetrized in a weighted mean,
using the individual uncertainties reported by pPXF to derive the
weights. The final kinematic uncertainties are computed as the errors
in the weighted mean values.

\begin{figure}
\centering \includegraphics[trim=20.pt 40.pt 40.pt 40.pt, clip, width=0.5\textwidth]{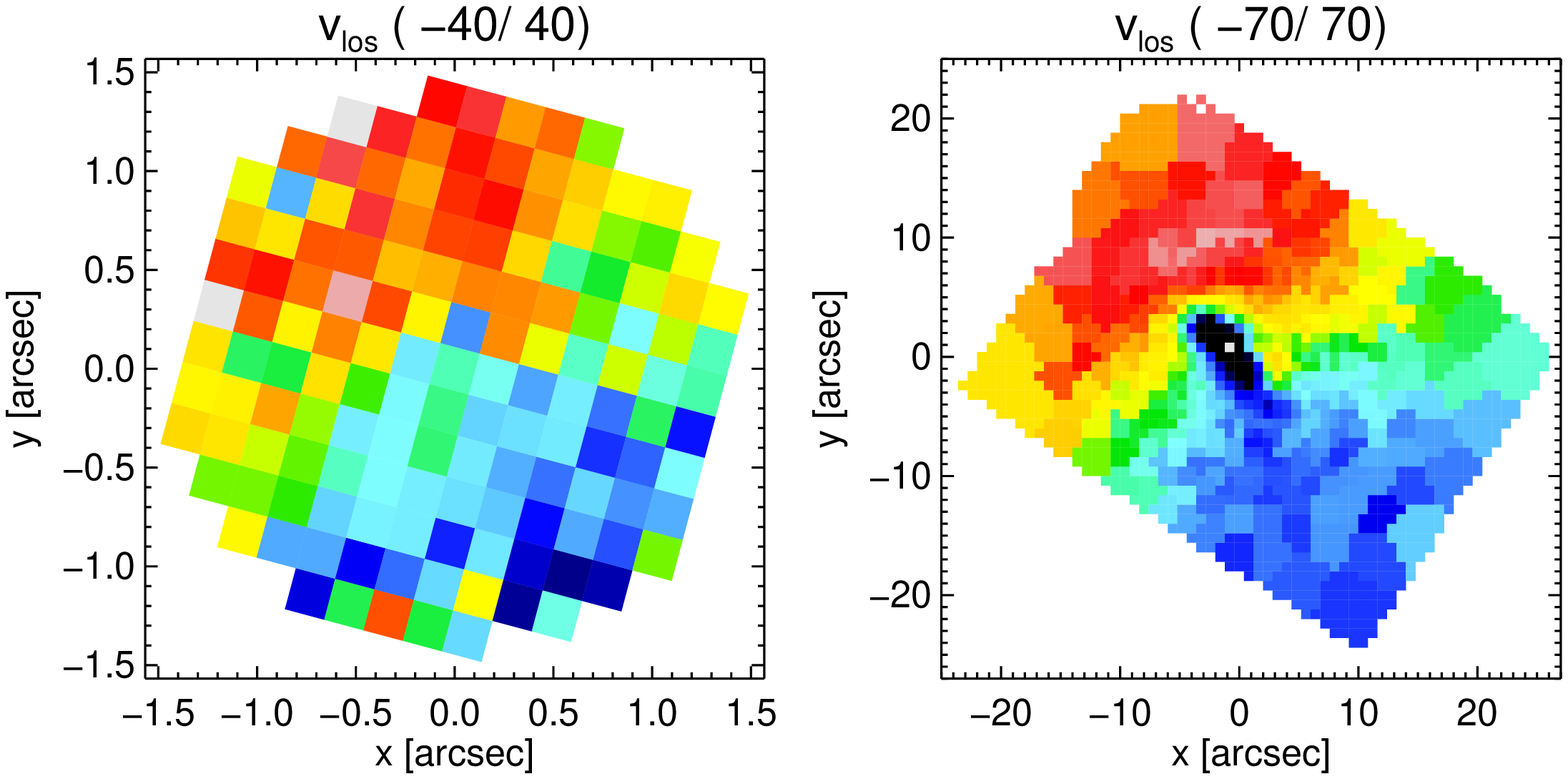}
\caption{Comparison of stellar velocity fields observed with the NIFS
  IFU (left) and SAURON IFU (right; Dumas et al.\ 2007). The
  AO-assisted NIFS data are consistent with the larger scale velocity
  map, but probe a spatial region inaccessible to the optical data
  because the AGN emission precludes measurements of the central
  stellar dynamics (the black region in the right panel). In both
  figures North is up and East is left.}
\label{fig:nifs_sau}
\vspace{0.2cm}
\end{figure}

Figure~\ref{fig:nifs_sau} shows the line-of-sight stellar velocity
fields from NIFS (left) and SAURON (right). The latter data are from
\citet{dumas_etal_07} and show a similar large scale rotational
velocity to that seen in the inner 3\arcsec$\times$3\arcsec\ region
observed with NIFS. Although the large scale line-of-sight velocity
fields in both panels show the same global rotational velocity, the
central region of the SAURON velocity field within ~5\arcsec $\times$
5\arcsec\ (right) is dominated by AGN continuum. Therefore, the
innermost parts allow no velocity measurement, and appear as a black
oval region in the image. Thus, the SAURON data do not provide
constraints on the BH mass, although they could provide constraints on
the mass-to-light ratio ($\Upsilon$). However, we do not include the
SAURON data in our modeling, relying instead on the long-slit
kinematics obtained from MMT and KPNO to provide the constraints on
$\Upsilon$ at large radii. This is partly a pragmatic choice since the
use of IFU data on both small and large scales would significantly
increase the total number of kinematic constraints that would need to
be fitted and, therefore, would require a significantly larger orbit
library than we use here, making the problem significantly more
computationally expensive than we are able to handle.

\subsection{Surface Brightness Profile}

We calculate the $H$-band surface brightness profile using our nested
series of images to probe different spatial scales. For large radii
($5\arcsec-90\arcsec$), we use the photometrically calibrated OSUBSGS
images. At intermediate radii ($1\arcsec-11\farcs25$), we use the ACS
optical image, allowing a significant region of overlap with the
OSUBSGS data. The ACS flux level is then rescaled to provide the best
match to the $H$-band data in the overlap region. Our use of the
optical data could potentially bias the surface brightness profile we
derive if there are significant variations in extinction or stellar
population at radii of $1\arcsec-5\arcsec$. However, there are no
independent indications of stellar population changes in that part of
NGC~4151, and the radial profile of the ACS image is relatively smooth
at those distances. Those radii are also well isolated from the PSF of
the AGN in the ACS image.

In the central regions, we derive a stellar flux profile directly from
the NIFS datacube. One of the outputs of the pPXF modeling is the
derived amplitude of the stellar template that best matches the galaxy
data. As our template spectra are continuum-normalized, these
amplitudes constitute relative measurements of the stellar flux in
each spaxel, calculated in a way that naturally omits the AGN
contribution to the light profile. The unbinned data are used to probe
radii of 0$\farcs04-1\farcs2$ (excluding the AGN-dominated central
spaxel). The overall NIFS stellar profile is then scaled in amplitude
to match the overlapping ACS data from larger radii.

We then fit the full $H$-band surface brightness profile using the
Multi-Gaussian Expansion (MGE) package
\citep{emsellem_etal_94,Cappellari02}.  The MGE routine\footnote{Also
  available at \url{http://purl.org/cappellari}} fits the smallest
necessary number of nested 2D Gaussians with a common center and
radially varies their amplitudes and axis ratios to obtain the best
possible smooth fit to the observed 2D surface brightness
distribution, taking account of the PSF. The deprojection of a 2D
surface brightness distribution to derive the 3D luminosity
distribution suffers from the well known degeneracy arising from
``konus densities'' that are invisible in projection
\citep{gerhard_binney_96,kochanek_rybicki_96}. A number of different
methods have been used to derive the 3D luminosity density, none of
which guarantee a unique deprojection. The MGE method has the
advantage that, once an angle of inclination is assumed, each 2D
Gaussian has a unique 3D luminosity density whose potential is easily
computed and which is positive everywhere \citep{Cappellari02}. The
code outputs the amplitude (total counts), width in pixels
($\sigma_{pix}$), and observed axis ratio ($q$) for each Gaussian. We
convert those values to a central surface brightness, $I_{H}$, for
each Gaussian as:
\begin{equation}
  \log_{10} I_{H} {\rm (L_{\odot,H}/pc^{2})}= 0.4 (\mu_{\odot,H} - \mu_{H}),
\end{equation}
where $\mu_{\odot,H}$ is the $H$-band surface brightness of the Sun at a
distance where 1~pc$^{2}$ subtends 1~arcsec$^{2}$. For an absolute
$H$-band magnitude of 3.32$\pm$0.03~mag \citep{binney_merrifield_98},
this gives $\mu_{\odot,H}=22.89$~mag~arcsec$^{-2}$.  The surface
brightness is computed from the MGE fits as:
\begin{equation}
  \mu_{H} = m_{0} - 2.5 \log_{10} (\frac{C_{tot}}{2\pi q \sigma_{pix}^{2}}) + 5
  \log_{10} S_{pix} - A_{H},
\end{equation}
where $m_{0}$ is the OSUBSGS $H$-band image zero-point of 22.2~mag,
$C_{tot}$ is the total counts, $S_{pix}$ is the plate scale of
$1\farcs5$~pixel$^{-1}$, and the $H$-band extinction is
$A_{H}$=0.016~mag \cite[][retrieved from the NASA/IPAC Extragalactic
Database]{Schlegel98}. For the MGE modelling, a 1-component PSF was
used, having an effective width of $\sigma=0.05$.

\begin{figure}
\centering \includegraphics[trim=0.pt 0.pt 0.pt 0.pt, width=0.4\textwidth]{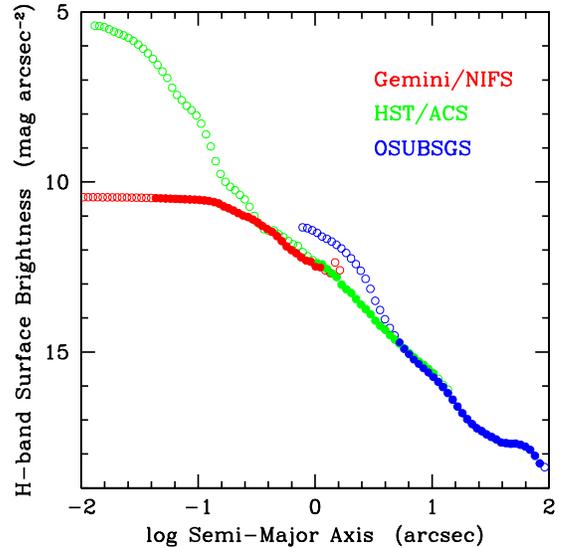}
\caption{Composite $H$-band radial surface brightness profile of the
  bulge, consisting of data from Gemini/NIFS at small radii (colored
  red in the online version), HST/ACS at intermediate radii (colored
  green in the online version), and OSUBSGS at large radii (colored
  blue in the online version). The OSUBSGS and HST/ACS surface
  brightness profiles both turn upwards at small radii because of the
  AGN. The NIFS profile comes from the amplitude of the stellar
  features in each spaxel and so naturally omits the AGN. The open
  points, which are contaminated by AGN flux, are not used in the MGE
  fit to the SB profile. In contrast to the data in
  Table~\ref{tab:MGE_SB}, the values plotted here have not been
  corrected for the PSF, in order to more clearly show the radial
  regions free from AGN contamination.}
\label{fig:SB_profile}
\vspace{0.2cm}
\end{figure}

The output of our MGE fit to the surface brightness distribution is
given in Table~\ref{tab:MGE_SB}. For each Gaussian in the MGE fit,
the first column gives an identifying number, the second column gives
the $H$-band luminosity density, the third column gives the standard
deviation $\sigma$ (in arcsec), and the fourth column gives the
flattening $q$ (ellipticity).  It is clear from the values of $q$ in
the third column that the bulge surface brightness distribution is
close to circular since most of the Gaussians have $q\sim 1$ (except
for the innermost and outermost Gaussians) suggesting that the bulge
is probably close to spherical. In Figure~\ref{fig:SB_profile}, we
show the composite radial surface brightness profile obtained by
patching together the three imaging datasets.

\begin{table}
\caption{MGE Fit to Surface brightness distribution}
\begin{centering}
\begin{tabular}{lccc}\hline
Gaussian & Surface Density, $I_H$ & Gaussian $\sigma$ & Axis Ratio $q$\\
Number & ($L_{\odot,H}$/pc$^2$)& (\arcsec)  & \\
(1) & (2) & (3) & (4)\\
\hline\\
1&   9.3963E+04      &     0.126    &  0.7906\\
2&   1.3200E+05      &    0.232     & 1.0000\\
3&   7.8100E+04      &     0.549    & 1.0000\\
4&   2.0951E+04      &      1.421   & 1.0000\\
5&   6.7191E+03      &      2.799   & 1.0000\\
6&   3.7009E+03      &     6.703    & 1.0000\\
7&   6.8485E+02      &    14.513   &  0.9550\\
8&   3.1906E+02      &    64.987   &   0.7052\\
\hline
\end{tabular}
\end{centering}
\label{tab:MGE_SB}
\vspace{1cm}
\end{table}

\subsection{Mass-to-Light Ratio}\label{sec:mtol}

To provide a rough constraint on the relationship between the
luminosity distribution and the stellar mass distribution, we combine
optical imaging from SDSS with the OSUBSGS $H$-band image. We
construct ($g-i$) and ($i-H$) color maps of NGC~4151, finding the
bulge colors of ($g-i$)=1.1$\pm$0.1~mag, and ($i-H$)=2.4$\pm$0.2~mag
to be quite uniform across the bulge (aside from the central few
arcseconds, where the AGN dominates the
emission). \citet{zibetti_etal_09} calculated $H$-band mass-to-light
ratios ($\Upsilon_H$) in ($g-i$)-($i-H$) color-space, based on a Monte
Carlo analysis of \citet{bruzual03} stellar population synthesis
models. Examining our measured colors in the \citet{zibetti_etal_09}
color-space indicates that $\Upsilon_H = 0.4\pm0.2~\msun/\lsun$. Our
initial exploration with dynamical models spanning the entire range
$0.2~\msun/\lsun < \Upsilon_H < 0.6~\msun/\lsun$ showed that the best
fit values were always in the range $0.2~\msun/\lsun < \Upsilon_H <
0.425~\msun/\lsun$, hence this range of values is used in the models
below.

\subsection{Morphological Classification}

\begin{figure*}
\centering \includegraphics[trim=0.pt 0.pt 0.pt 0.pt, width=0.3\textwidth]{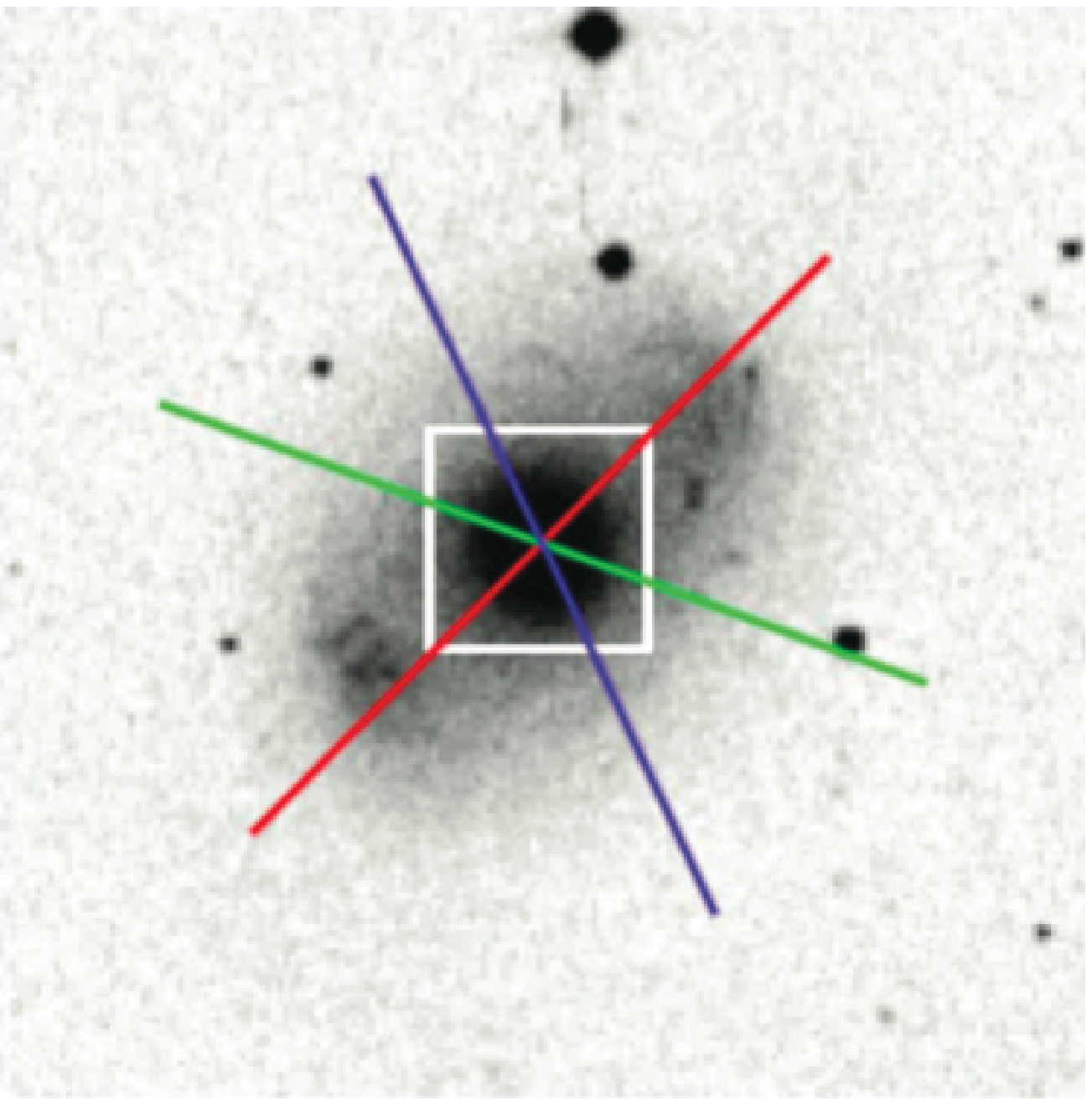}
\centering \includegraphics[trim=0.pt 0.pt 0.pt 0.pt, width=0.3\textwidth]{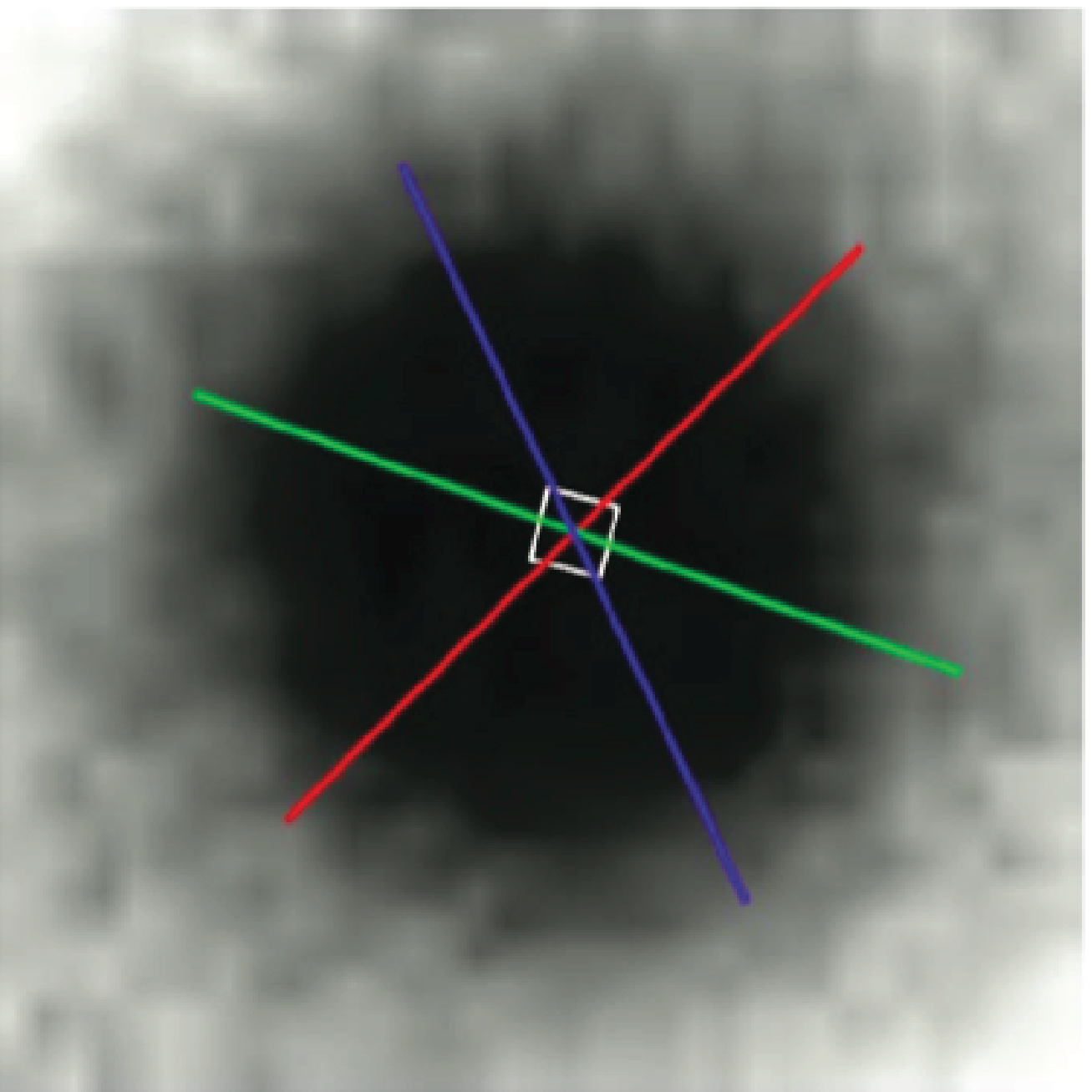}
\centering \includegraphics[trim=0.pt 0.pt 0.pt 0.pt, width=0.35\textwidth]{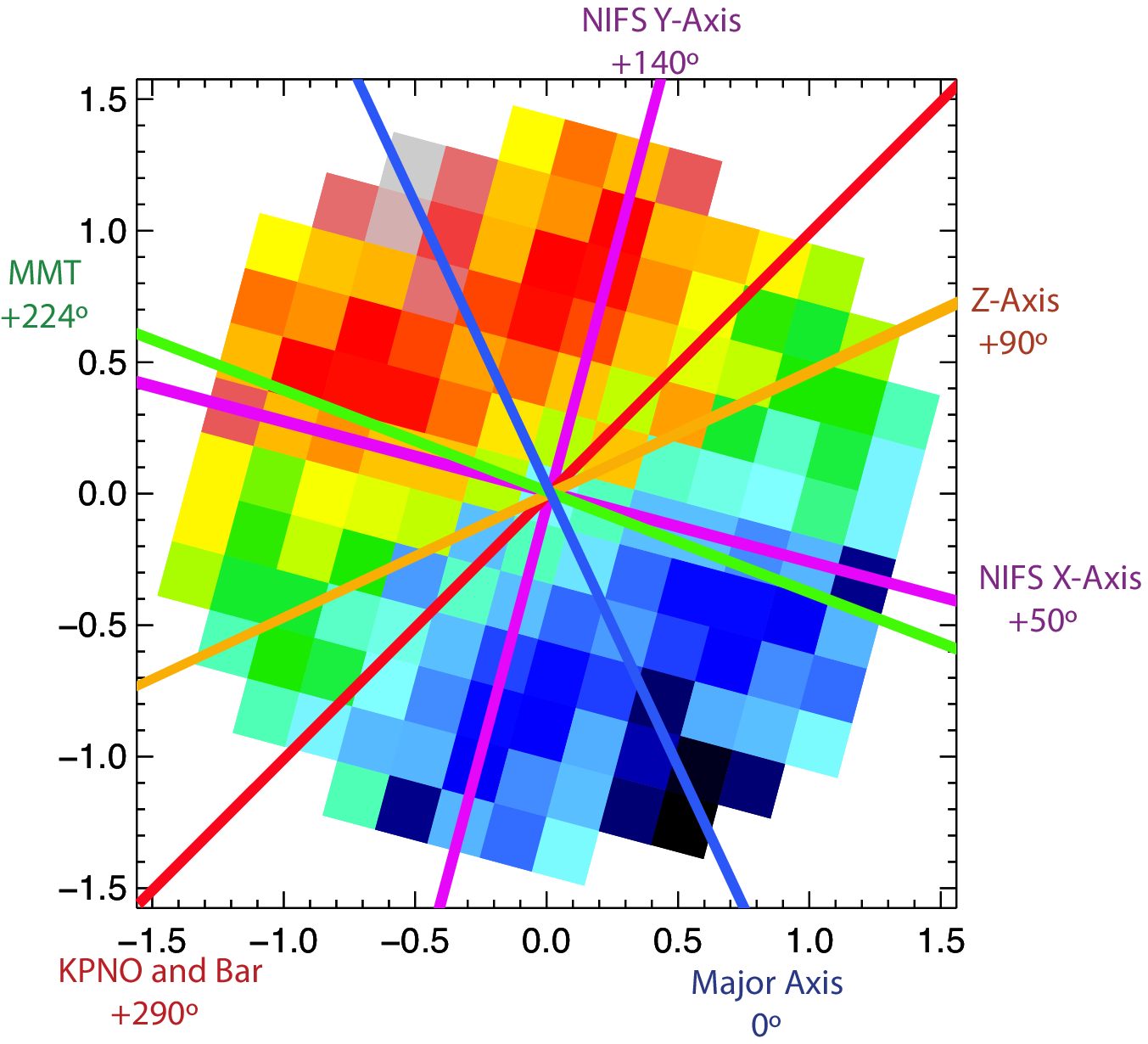}
\caption{Orientation of long-slits and NIFS field to kinematic and
  photometric axes of the galaxy. Left: Optical image of NGC~4151
  (from POSS). North is at top, East at left. The field-of-view is
  5\arcmin$\times$5\arcmin. The white box is 1\arcmin$\times$1\arcmin
  and is shown zoomed-in in the middle panel. The red line shows the
  orientation of the KPNO slit (2\arcsec wide), the green line shows
  the orientation of the MMT slit (1\arcsec wide), the blue line shows
  the orientation of the major axis of the bulge as inferred from the
  symmetry axis of the velocity field (see right panel). Middle: the
  nuclear region 1\arcmin$\times$1\arcmin; the grey box shows the size
  and orientation of the NIFS field. Right: Symmetrized NIFS velocity
  field showing inferred kinematic major axis (blue line), minor axis
  (Z; ochre line). The angles of the slits and NIFS instrument are
  given relative to the kinematic major axis.}
\label{fig:slit_orient}
\vspace{0.2cm}
\end{figure*}

NGC~4151 was classified by \citet{devaucoulers_etal_91} as
(R')SAB(rs)ab: a disk galaxy with a pseudo-outer ring (R'), a
``mixed'' bar (AB) (i.e., possibly barred but is not strongly so),
with mixed spiral/inner ring classification (rs), and an intermediate
to large bulge (ab). Figure~\ref{fig:slit_orient}~(left) shows the
Palomar Observatory Sky Survey (POSS) image of NGC~4151 in the red
optical filter\footnote{From the Digitized Sky Survey; retrieved via
  the NASA/IPAC Extragalactic Database:
  http://ned.ipac.caltech.edu/}. In part because of the very circular
bulge, the classification of this galaxy as barred has been questioned
by authors as far back as \citet{davies_73}, who speculated that the
elongation was not a bar, but was in fact made up of material that was
ejected from the nucleus in a former explosive event. However, we
  believe that NGC~4151 may be an example of a galaxy containing a
  ``barlens'', an oval-shaped structure that is likely to be the
  vertically thick aspect of a bar seen from a particular viewing
  angle \citep{laurikainen11,athanassoula14}.

In Paper~I, we assumed that the kinematic major axis is aligned with
the bar, since spectral information was only available along two
long-slits. The KPNO slit (represented by the red line in
Fig.~\ref{fig:slit_orient}~left) was oriented along the bar at a
position angle of 135$\deg$ \citep{ferrarese_etal_01}. The MMT slit
(green line in Fig.~\ref{fig:slit_orient}~left) was placed at a
position angle of 69$\deg$.  The middle panel of
Figure~\ref{fig:slit_orient} shows the region within the white box
from the left panel. The grey box in the middle panel represents the
orientation and shape of the NIFS field of view. It is quite clear
from a comparison of the NIFS velocity field and the SAURON velocity
field in Fig.~\ref{fig:nifs_sau}, that the rotation axis of the disk
and bulge are not perpendicular to the bar as was assumed in
Paper~I. This is an important difference between the models presented
here and the models presented in Paper~I.

\subsection{Distance to NGC~4151}

To first order, the BH mass derived by the stellar dynamical modeling
depends linearly on the assumed distance to the galaxy\footnote{This
  arises simply from the fact that in virial systems, $M\sim v^2/R$
  where R is the linear radius within which the velocity (or velocity
  dispersion) is measured, and the conversion between angular and
  linear radii is inversely proportional to distance in the nearby
  universe.}. For consistency with the analysis in Paper~I, we adopt a
distance for NGC~4151 of 13.9~Mpc \cite[based on the recession
velocity of 998~\kms\ measured by][]{pedlar92}. However, as noted in
Paper~I, the actual distance to the galaxy is rather uncertain, with
published estimates spanning a range of 10~Mpc to 30~Mpc. Recently,
\citet{Tully09} compiled a group-averaged Tully-Fisher distance of
11.2$\pm$1.1~Mpc in the Extragalactic Distance
Database\footnote{\url{http://edd.ifa.hawaii.edu}}. However, that
group distance is based on just four galaxies, which have a standard
deviation in their distance moduli of more than four times the
individual distance modulus errors. Furthermore, it is unclear whether
the contribution to the group average distance from NGC~4151 (of
$\sim$4~Mpc) accounts for the AGN contribution to the galaxy
brightness. Thus, we cannot regard the \citet{Tully09} value as being
any more reliable than other distances, and until a more accurate
distance can be obtained, we continue to use the value of 13.9~Mpc.

\section{Stellar Dynamical Modeling}
\label{sec:dynamics}

We apply what is now the standard stellar dynamical modeling approach
of orbit superposition
\citep[e.g][]{schwarzschild_79,vandermarel_etal_98,cretton_etal_99,gebhardt_etal_03,valluri_etal_04}
to fit the observed LOSVDs and luminosity distribution of
NGC~4151. The orbit superposition (or ``Schwarzschild'') method of
dynamical modeling uses line-of-sight kinematical and imaging data to
constrain the mass distributions of external galaxies. The
implementation of the code used here differs from the method detailed
in \citet[][hereafter VME04]{valluri_etal_04} in two small aspects. It
is (a) designed to fit generalized axisymmetric models that are
obtained from direct deprojection of the surface brightness profiles
of the individual galaxies using a MGE code, following the method
outlined in \citet{Cappellari02}; (b) modified to fit integral field
kinematics in addition to long-slit kinematics.

\subsection{Model Setup}

We restrict our models to the surface brightness distribution within
50\arcsec --- i.e., the bulge region --- using the MGE decomposition
from Table~\ref{tab:MGE_SB}. For each dynamical model, we assume a
constant $H$-band mass-to-light ratio ($\Upsilon_H$) for the
stars\footnote{We do not include a dark matter halo in our model since
  this would require a lower $\Upsilon$ value for the stars and the
  best-fit model values of $\Upsilon_H$ described below are completely
  consistent with the expectations derived from the stellar photometry
  in Sec.~\ref{sec:mtol}. Even if we were to include such a halo,
    its contribution to the inner 1\farcs5 region, where the NIFS
    kinematics exist, would likely be small. As we will show, the NIFS
    data entirely determine our final BH mass, justifying our decision
    to not include a dark matter halo.}, with values varying from
$0.2~\msun/\lsun - 0.425~\msun/\lsun$. The resulting density profile
is then converted to a gravitational potential using an axisymmetric
multipole expansion scheme \citep{BT}. The gravitational potential of
a point mass ($\mbh$), representing the central BH, is then
added. Models are constructed for every pair of parameters ($\mbh$,
$\Upsilon_H$), building an orbit library for each combination.

The large-scale \ion{H}{1} disk of NGC~4151 is inclined to the line of
sight at $i=20\deg-25\deg$ \citep{simkin_75}. However, since the bulge
appears very close to circular in projection (see
Fig.~\ref{fig:slit_orient} and Tab.~\ref{tab:MGE_SB}), it is probably
nearly spherical or being viewed close to face on. Thus, we run models
on a 2D grid of ($\mbh, \Upsilon_H$) values for three different
inclination angles $i=90\deg$ (edge-on), $i=60\deg$ and $i=23\deg$
(close to face-on). We also run a limited set of models for which the
value of $\mbh$ is held fixed while the inclination angle of the model
is varied with values $i= 15\deg, 30\deg, 50\deg$, and $75\deg$, in an
effort to set constraints on the inclination of the bulge to the
line-of-sight.
  
For inclination angles of $i=90\deg$ and $23\deg$, we run orbit
libraries for 18 BH masses between $1\times10^4~\msun - 2.6\times
10^8~\msun$, using a baseline $\Upsilon_H=0.3~\msun/\lsun$. These
models are then scaled to create libraries for 18 different
$\Upsilon_H$ values, producing a total of 324 models for each
inclination angle. We also run orbit libraries for $i=60\deg$, using
10 BH masses across the same mass range and 18 $\Upsilon_H$ values for
each $\mbh$ (180 models). The range of BH mass values was set by
initial experimentation with a coarse grid on a wide range of
values. The lowest two values of $\mbh$ are set to 10$^4~\msun$ and
10$^7~\msun$, both of which consistently give poor fits to the
data. The upper limit of $\mbh$ was raised until it was clear that
models once again gave very poor fits to the data.

The combined gravitational potential of the bulge and the BH is used
to integrate a large library of $N_o$ stellar orbits selected on a
grid with $N_E$ energy values, $N_A$ angular momentum values ($L_z$)
at each energy, and $N_{I3}$ pseudo third integral values ($I_3$) at
each angular momentum value (see VME04 for details). For instance,
models with $N_o= 9936$ used $N_E=46$, $N_A=24$, $N_{I3}=9$.  Each
orbit is integrated for 100 orbital periods and its average
``observed'' contribution to each of the ``apertures'' in which
kinematical data are available is stored. The orbital kinematics are
subjected to the same type of instrumental effects as the real data
(i.e. PSF convolution, pixel binning).

Self-consistent orbit-superposition models are constructed by linearly
co-adding orbits in each library to simultaneously fit several
elements within the observed apertures (imaging and spectroscopic):
the 3D mass distribution (for each value of $\Upsilon_H$), the
projected surface density distribution, and the LOSVDs. The fits use a
non-negative least squares optimization algorithm
\citep[NNLS;][]{NNLS}, which gives the weighted superposition of the
orbits that best reproduce both the self-consistency constraints
(i.e., the 3D mass distribution and surface brightness distribution)
and the observed kinematical constraints ($V_{\rm los}, \sigma_{\rm
  los}, h_3, h_4$).  

VME04 showed that as the ratio of orbits to constraints
decreases, the error bars decrease artificially. In the majority of
the models presented here we have 1240 constraints that are fitted by
a library of 9936 orbits. This size of orbit library gives an
orbits-to-constraints ratio of 8 which is only slightly above the
ratio (5) at which VME04 find that solutions begin to be biased. In
Section~\ref{sec:robust_size} we carry out a limited exploration with a
50\% larger orbit library and find that our results are unchanged. 

Smoothing (or ``regularization'') of the orbital
solutions is typically performed in this type of analysis, and a
variety of methods have been employed to that end.  Regularization
mainly helps to reduce the sensitivity of the solutions to noise in
the data by requiring a smoother sampling of the orbit libraries used
in the solution. Following \citet{cretton_etal_99}, we run a trial
series of models with a regularization scheme that requires local
smoothing in phase space. We find that for small smoothing parameters
($\lambda = 0.1$) the resultant best-fit $\mbh$ values do not differ
from the models without regularization. However, the $\chi^2$ value
for the best-fit model is larger and the error bars on $\mbh$ are
smaller. VME04 showed that it is impossible (without computationally
expensive approaches such as generalized cross-validation) to
determine the ideal regularization parameter. Not regularizing the
solutions can yield large error bars, but choosing too large a value
of the smoothing parameter can introduce a bias in the best-fit value
of $\mbh$. In this paper, we present models without regularization,
since imposing regularization constraints significantly increases the
total number of constraints and therefore decreases even further the
orbits-to-constraints ratio. For the libraries used here, models
without regularization provide the most conservative error bars on the
estimated value of $\mbh$.

\begin{figure}
\centering \includegraphics[trim=0.pt 0.pt 0.pt 0.pt, width=0.4\textwidth]{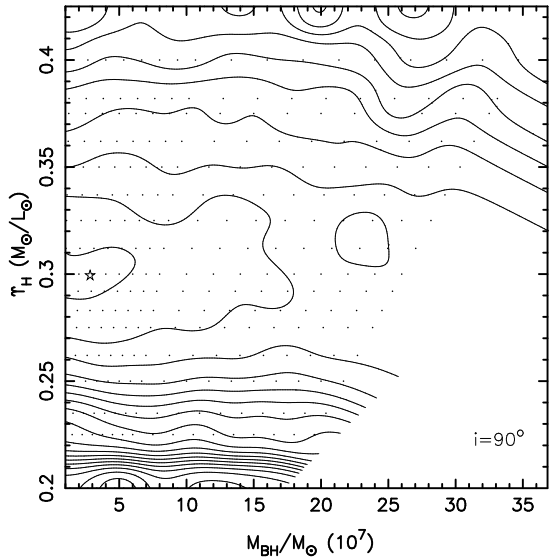}
\caption{Contours of constant {\em fractional error} in enclosed mass
  and aperture surface brightness (self-consistency constraints) in
  the plane of parameters $\Upsilon_H$ and $\mbh$ for models with
  inclination to the line of sight $i=90\deg$. The star marks the
  location of the model with 0.27\% fractional error per
  constraint. Contours are spaced at intervals of 0.1\% fractional
  error per constraint.  }
\label{fig:mass_err}
\vspace{0.2cm}
\end{figure}

\subsection{Model Self-Consistency Constraints}

Since the mass and surface mass density constraints are {\em
  self-consistency constraints} rather than observational constraints,
they do not have observational errors associated with them. There are
two ways in which the absence of observational errors are handled in
the literature. \citet{cretton_etal_99} and VME04 include
self-consistency constraints in the NNLS minimization problem with
errors corresponding to predefined level of relative accuracy in the
fit (e.g. 5\% -10\%), while others
\citep{gebhardt_etal_03,vandenbosch_etal_08} solve the orbit
superposition problem by requiring the self-consistency constraints
(mass and surface brightness) to be fitted as separate as linear
(in)equality constraints with predetermined absolute accuracy.
\citet{vandenbosch_etal_08} state that they require the
self-consistency constraints to be fitted to an accuracy of 0.02. We
also attempted to fit self-consistency constraints as linear
equalities but find that that since the values of the mass constraints
vary by 5 orders of magnitude (from $10^{-4}$ to 10), using a fixed
numerical accuracy fails to fit the mass constraints at small radii
(where the logarithmic radial bins have the smallest mass values per
bin and where the fit to mass constraints most strongly affects the
estimated $\mbh$).

For the models presented here we include the self-consistency
constraints in the optimization problem and require the optimization
problem to fit each of the 472 self-consistency constraints (280 mass
constraints and 192 aperture surface brightness constraints) with an
error corresponding to a relative accuracy-per-constraint of 5\%. For
our preferred model with $i=90\deg$, the fractional accuracy per
self-consistency constraint varies from 0.27\% (for the best fit
model) to 2.2\% for the worst fit model, i.e., always better than our
required accuracy. Figure~\ref{fig:mass_err} shows contours of
constant fractional error in the fit to the self-consistency
constraints (mass and surface brightness in apertures) over the 2D
parameter space ($\mbh$, $\Upsilon_H$) for models with $i=90\deg$. The
star is the location of the model with 0.27\% accuracy per
self-consistency constraint, and the contours are spaced at intervals
of 0.1\% relative accuracy per constraint.

In previous applications of our Schwarzschild modeling code
\cite[VME04,][Paper~I]{valluri_etal_05}, the error in the fit to the
self-consistency constraints was essentially random over the entire 2D
parameter space ($\mbh, \Upsilon$). Although the self-consistency
constraints were always included in the $\chi^2$, they contributed so
little to the total $\chi^2$ that the best fit solution was always
driven by the fit to the kinematical constraints. We will show that
this is not true for our current NGC~4151 dataset. Despite the fact
that the fits to the 3D mass distribution and surface brightness
distribution are very good, because the fit
vary systematically rather than randomly across the parameter space,
the best fit solution is altered by the decision of whether or not to
include the fit to the self-consistency constraints in the $\chi^2$.

Based on the arguments above, we compute the $\chi^2$ in two different
ways: (1) the total $\chiall$ includes all the constraints in the
NNLS optimization problem: 280 mass constraints, 192 surface
brightness constraints, and 768 kinematical constraints (giving a
total of $N_c=1240$ constraints); (2) $\chikin$ which only considers
the quality of the fit to $N_c=768$ kinematical constraints ($V_{\rm
  los}, \sigma_{\rm los}, h_3,$ and $h_4$ in 192 apertures: 136 NIFS
spaxels, 15 KPNO apertures, and 41 MMT apertures). For each of the
three assumed angles of inclination for the bulge, we construct models
by fitting these constraints using an orbit library consisting of
$N_o= 9936$~orbits.  For one inclination angle ($i=90\deg$) we do a
limited exploration with a library consisting of $N_o= 15092$
orbits. After generating orbit libraries for a grid of parameters ($\mbh,
\Upsilon_H$) the best fit parameters are determined by the minimum in
the 2D $\chi^2$-contour plot.

\section{Results}
\label{sec:results}

In this section, we present the results of our modeling. In
Section~\ref{sec:define_chi2} we describe results for inclination
angles $i=90\deg$, $i=60\deg$ and $i=23\deg$, showing 2D contour plots
of $\chiall$ and $\chikin$ using a full 2D grid of models for each of
these inclination angles. In Section~\ref{sec:inclination} we examine
the dependence on inclination angle for 4 additional models using a
single $\mbh$. In Section~\ref{sec:robust} we explore the robustness
of our solutions by examining their sensitivity to the size of the
orbit library and to the effect of restricting the set of kinematic
constraints to those from the nuclear region only. We also present
some of the best-fit solutions and show how well different pairs of
$\mbh, \Upsilon_H$ fit both the 1D and 2D kinematical data. In Section
5.4, we discuss our results in the context of the \msigma relation.

\subsection{Dependence on Definition of $\chi^2$}
\label{sec:define_chi2}

\begin{figure}
\centering \includegraphics[trim=0.pt 0.pt 0.pt 0.pt, width=0.35\textwidth]{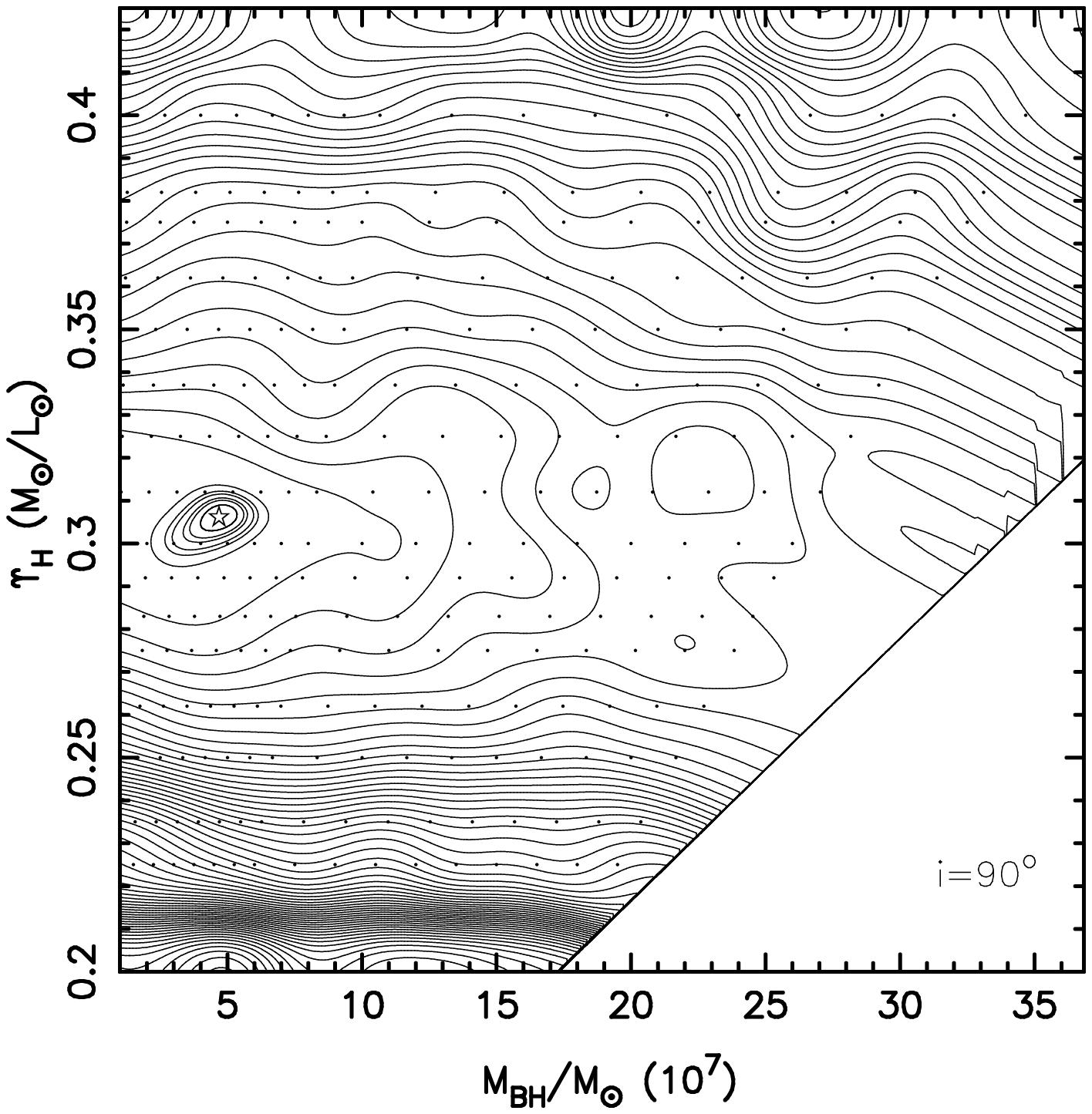}
\centering \includegraphics[trim=0.pt 0.pt 0.pt 0.pt, width=0.35\textwidth]{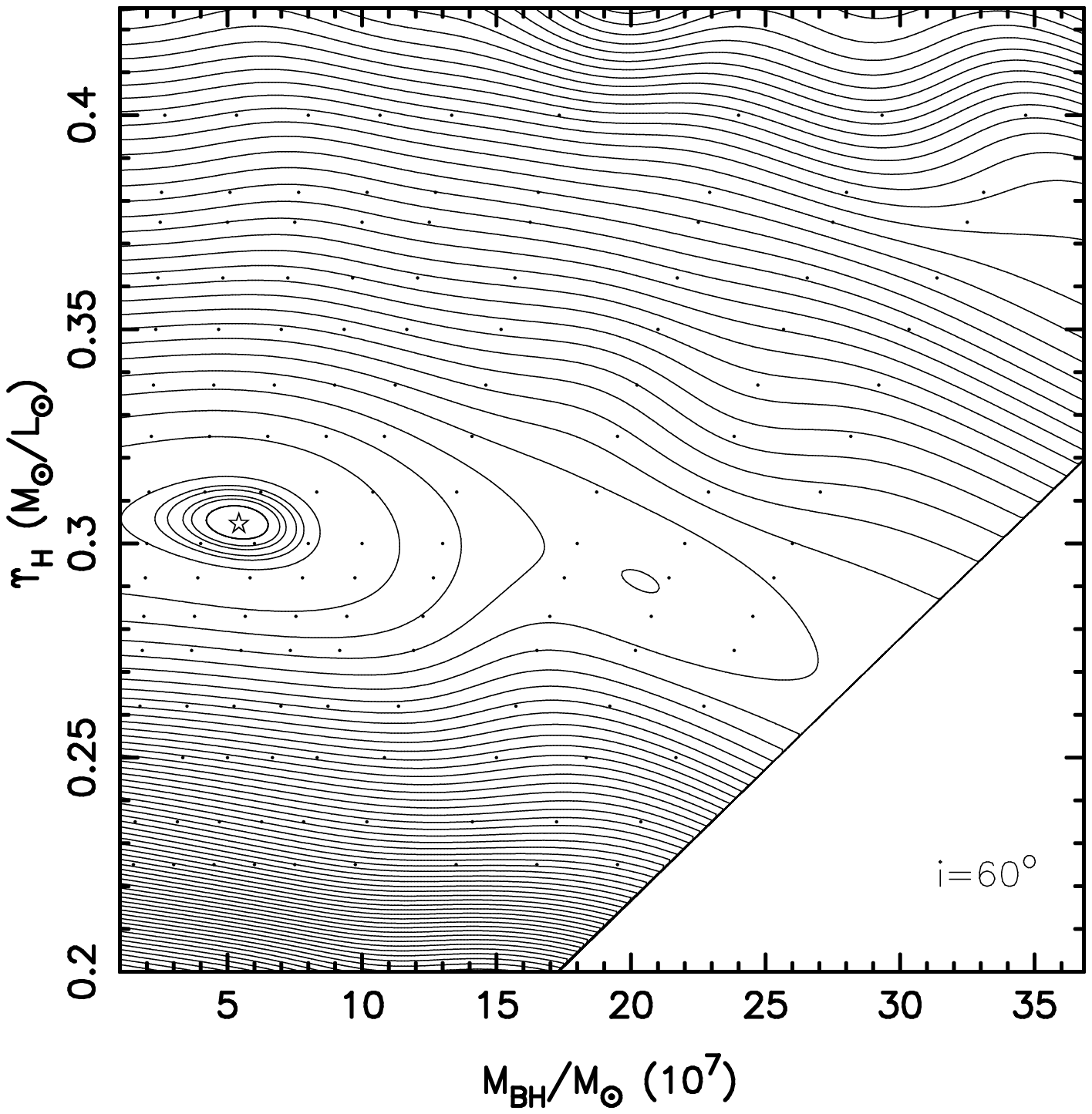}
\centering \includegraphics[trim=0.pt 0.pt 0.pt 0.pt, width=0.35\textwidth]{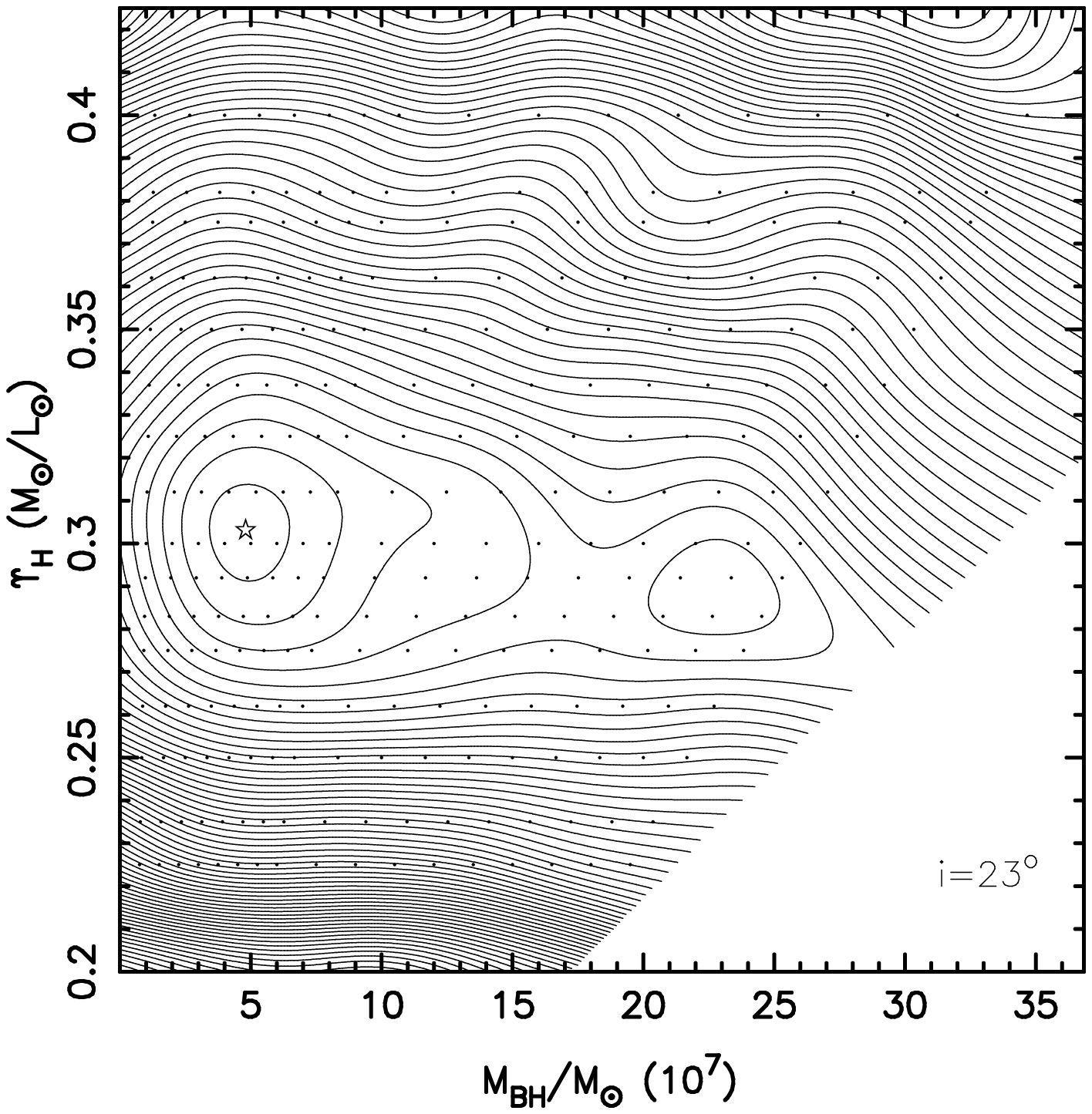}
\caption{Contours of constant $\chiall$ in the plane of model
  parameters $\Upsilon_H$ and $\mbh$ for models (each with $N_o=9936$)
  with inclination to the line of sight $i=90\deg$ (top), $i=60\deg$
  (middle), and $i=23\deg$ (bottom). The star in each plot marks the
  location of the best-fit values for \mbh and $\Upsilon_H$ (see
  Tab.~\ref{tab:bestfit} for details). The first six contours
  surrounding the star are the $ 1\sigma, 2\sigma, \ldots, 6\sigma$
  confidence intervals respectively. Subsequent contours are equally
  spaced between the 6th contour and the maximum $\chi^2$ value.  }
\label{fig:2dChi2}
\vspace{0.2cm}
\end{figure}

Figure~\ref{fig:2dChi2} shows 2D contour plots of $\chiall$ in the
plane of model parameters $\mbh$ and $\Upsilon_H$ for $i=90\deg$
(top), $i=60\deg$ (middle), and $i=23\deg$ (bottom). The grid of
points shows the model parameters ($\mbh, \Upsilon_H$) for which the
orbit libraries is constructed and the optimization problem is
solved. The first 6 contour levels in all the $\chi^2$ contour plots
that follow correspond to $\Delta \chi^2 = 2.3~(1\sigma$, 68.3\%
confidence), $\Delta \chi^2 = 4.61~(2\sigma$, 90\% confidence),
$\Delta \chi^2 = 6.17~ (3\sigma$, 95.4\% confidence), $\Delta \chi^2 =
9.21~(4\sigma$, 99\% confidence), $\Delta \chi^2 = 11.9~ (5\sigma$,
99.73\%), and $\Delta \chi^2 = 18.4~ (6\sigma$, 99.99\%), with
subsequent contours equally spaced in $\Delta \chi^2$.

For $i=90\deg$, the minimum value of $\chiall$ is obtained for $\mbh
=4.68\times 10^7~\msun$ and $\Upsilon_H = 0.304~\msun/\lsun$, with
$(\chiall)_{\rm min}=709.1$. For inclination angle of $i=60\deg$, the
best fit model has $\mbh = 5.42 \times 10^7~\msun$ and $\Upsilon_H =
0.305~\msun/\lsun$, with $(\chiall)_{\rm min}=767.5$. For inclination
angle of $i=23\deg$, the best fit model has $\mbh = 3.69 \times
10^7~\msun$ and $\Upsilon_H = 0.304~\msun/\lsun$, with $(\chiall)_{\rm
  min}=1050.8$.

\begin{figure}
\centering \includegraphics[trim=0.pt 0.pt 0.pt 0.pt, width=0.3\textwidth]{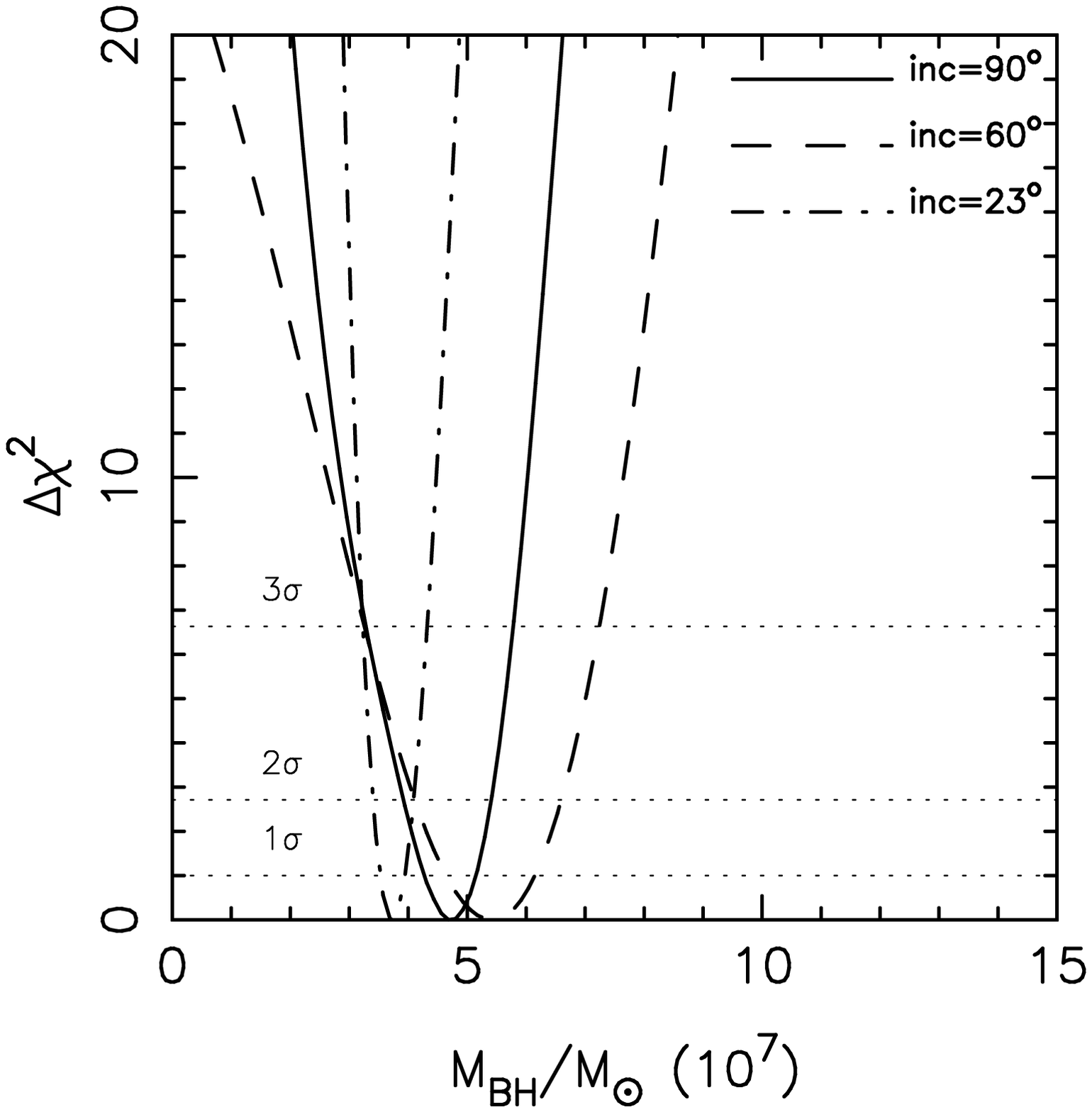}
\centering \includegraphics[trim=0.pt 0.pt 0.pt 0.pt, width=0.3\textwidth]{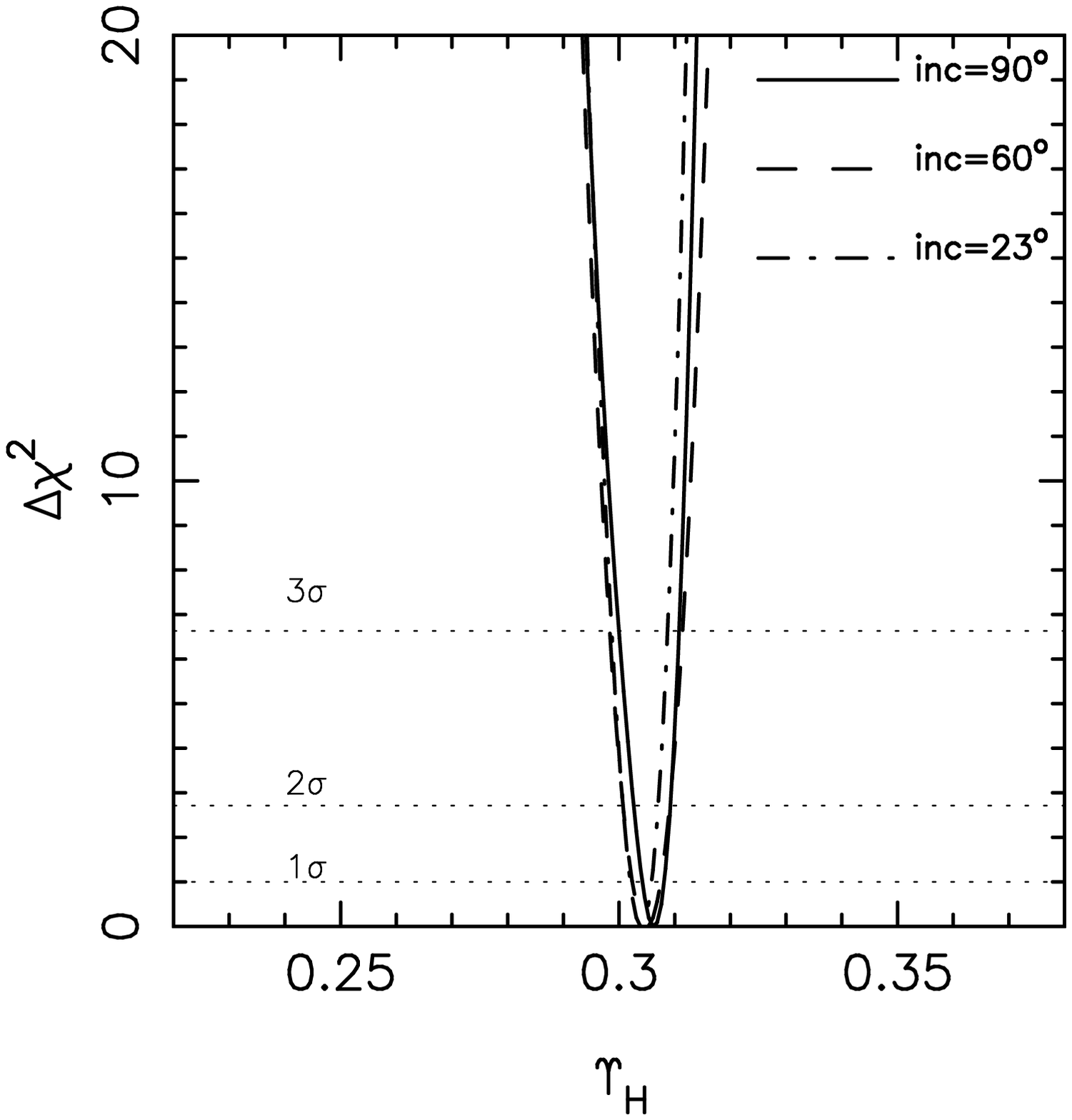}
\caption{$\Delta \chiall$ curves marginalized over $\Upsilon_H$ (top)
  and marginalized over $\mbh$ (bottom) in Fig.~\ref{fig:2dChi2}, for
  $i=90\deg$ (solid), $i=60\deg$ (dashed) $i=23\deg$ (dot-dashed). The
  thin horizontal dotted lines corresponds to $\Delta \chiall=1$
  (1$\sigma$, 68.5\% confidence interval), $\Delta \chiall=2.71$
  (2$\sigma$, 95\%) and $\Delta \chiall=6.63$ (3$\sigma$, 99\%). }
\label{fig:1dChi2}
\vspace{0.2cm}
\end{figure} 

Table~\ref{tab:bestfit} summarizes the results and includes minimum
$\chi^2$ values and the 1$\sigma$ and 3$\sigma$ error bars. The errors
on a given parameter are obtained by marginalizing the 2D $\chi^2_{\rm
  all}$ values over the other parameter. Figure~\ref{fig:1dChi2}~(top)
shows the marginalized 1D $\Delta \chi^2$ versus $\mbh$ for the three
different inclination angles, as indicated by the legends. The
horizontal dotted lines are at $\Delta \chi^2 = 1$ (1$\sigma$, 68.5\%
confidence interval for 1 degree-of-freedom), $\Delta \chi^2 = 2.71$
(2$\sigma$, 90\% confidence interval), and $\Delta \chi^2 = 6.63$
(3$\sigma$, 99\% confidence interval). The 1$\sigma$ and 3$\sigma$
errors on $\mbh$ and $\Upsilon_H$ are given in
Table~\ref{tab:bestfit}. Likewise, Figure~\ref{fig:1dChi2}~(bottom)
shows the marginalized 1D $\Delta \chi^2$ versus $\Upsilon_H$ for each
of the three inclinations of models to the line-of-sight.

It is important to note that with the standard methods used to derive
the velocity profiles (VPs) from the spectra, the errors in the
Gauss-Hermite (GH) coefficients are not independent. The errors
associated with the even moments ($\sigma$ and $h_4$) are correlated,
and errors associated with the odd moments ($V$ and $h_3$) are also
correlated \cite[see,
e.g.,][]{joseph_etal_01,cappellari_emsellem_04}. \citet{houghton_etal_06}
point out that an important consequence of correlated errors in the GH
coefficients is that the errors obtained by comparing data to
dynamical models using the $\chi^2$ estimator are significantly
underestimated.\footnote{In order to properly take account of the
  co-variances in the GH moments, \citet{houghton_etal_06} have
  proposed decomposing the VPs into eigenfunctions (or
  ``eigen-VPs'').} In this paper we quote both 1-$\sigma$ and
3-$\sigma$ errors, and often consider the 3-$\sigma$ errors as better
representations of the true modeling errors.

\begin{table*}[ht]
\caption{Best-fit values of $\mbh$ and $\Upsilon_H$ and error bars}
\centering
\begin{tabular}{lcccccccc}\hline 
\multicolumn{1}{l}{$i$ }&
\multicolumn{1}{l}{$N_o$ }&
\multicolumn{1}{l}{$\chi^2_{\rm min}$ }&
\multicolumn{3}{c} {$\mbh (10^7~\msun$)}& 
\multicolumn{3}{c} {$\Upsilon_H$ ($\msun/\lsun$)} \\
\multicolumn{1}{l}{($\deg$)} &
\multicolumn{1}{c}{ } &
\multicolumn{1}{c}{ } &
\multicolumn{1}{c} {best-fit} &
\multicolumn{1}{c} {1$\sigma$} &
\multicolumn{1}{c} {3$\sigma$ }& 
\multicolumn{1}{c} {best-fit} &
\multicolumn{1}{c} {1$\sigma$ } &
\multicolumn{1}{c} {3$\sigma$}\\
(1) & (2) & (3) & (4) & (5) & (6) & (7) & (8) & (9)\\
\hline
\multicolumn{9}{c}{$\chi^2_{\rm all}$ includes mass, surface
  brightness, kinematics, $N_c=1240$}\\
\hline\\
90$\deg$& 9936 & 709.1& 4.68     & $^{+0.46}_{-0.41}$ & $^{+1.10}_{-1.40}$& 0.304 & $\pm0.002$ &$\pm0.005$\\
                  &          &              &                                 &                               &        &  &\\
60$\deg$& 9936 & 767.5 & 5.42      & $^{+0.73}_{-0.78}$ & $^{+1.82}_{-2.2}$& 0.305&$\pm0.003$  &$\pm0.006$\\
                 &            &             &                                  &                               &       &  &\\
23$\deg$ & 9936 & 1050.8 & 3.69  & $^{+0.25}_{-0.16}$   & $^{+0.62}_{-0.47}$& 0.304& $\pm0.003$  &$\pm0.005$\\
                 &            &             &                                  &                               &       &  &\\
\hline
\multicolumn{9}{c}{$\chi^2_{\rm kin}$ includes only kinematics, $N_c=768$}\\
\hline\\
90$\deg$& 9936 & 186.7 & 7.76      &  $^{+1.92}_{-1.89}$ & $^{+4.20}_{-4.09}$& 0.332 & $\pm0.008 $ &$\pm 0.022$\\
                  &          &              &                               &                               & &  &\\
60$\deg$ & 9936& 193.1 & 8.62      & $^{+1.22}_{-1.26}$& $^{+3.35}_{-3.10}$& 0.326&$\pm0.006$  &$\pm0.019$\\
                 &            &             &                               &                               & &  &\\
23$\deg$& 9936 & 348.2 &8.50    & $^{+1.01}_{-1.07}$& $^{+3.69}_{-3.25}$& 0.329& $\pm0.004$  &$\pm0.013$\\
                 &            &             &                               &                               & &  &\\    
\hline
\end{tabular}
\label{tab:bestfit}
\vspace{0.5cm}
\end{table*}

The best fit solutions for both $\mbh$ and $\Upsilon_H$ for the three
different inclination angles are consistent with each other within
3$\sigma$. However $(\chiall)_{\rm min}$ is the smallest for the
edge-on model ($i=90\deg$), increases for $i=60\deg$ and is
significantly larger for $i=23\deg$ implying that it may be possible
to constrain the inclination of the bulge (see
Section~\ref{sec:inclination}). Marginalizing $\chiall$ over the three
values of inclination gives a best fit $\mbh \sim 5 \times10^7 \msun$
and $\Upsilon_H \sim 0.30\msun/\lsun$.

\begin{figure}
\centering \includegraphics[trim=0.pt 0.pt 0.pt 0.pt, width=0.35\textwidth]{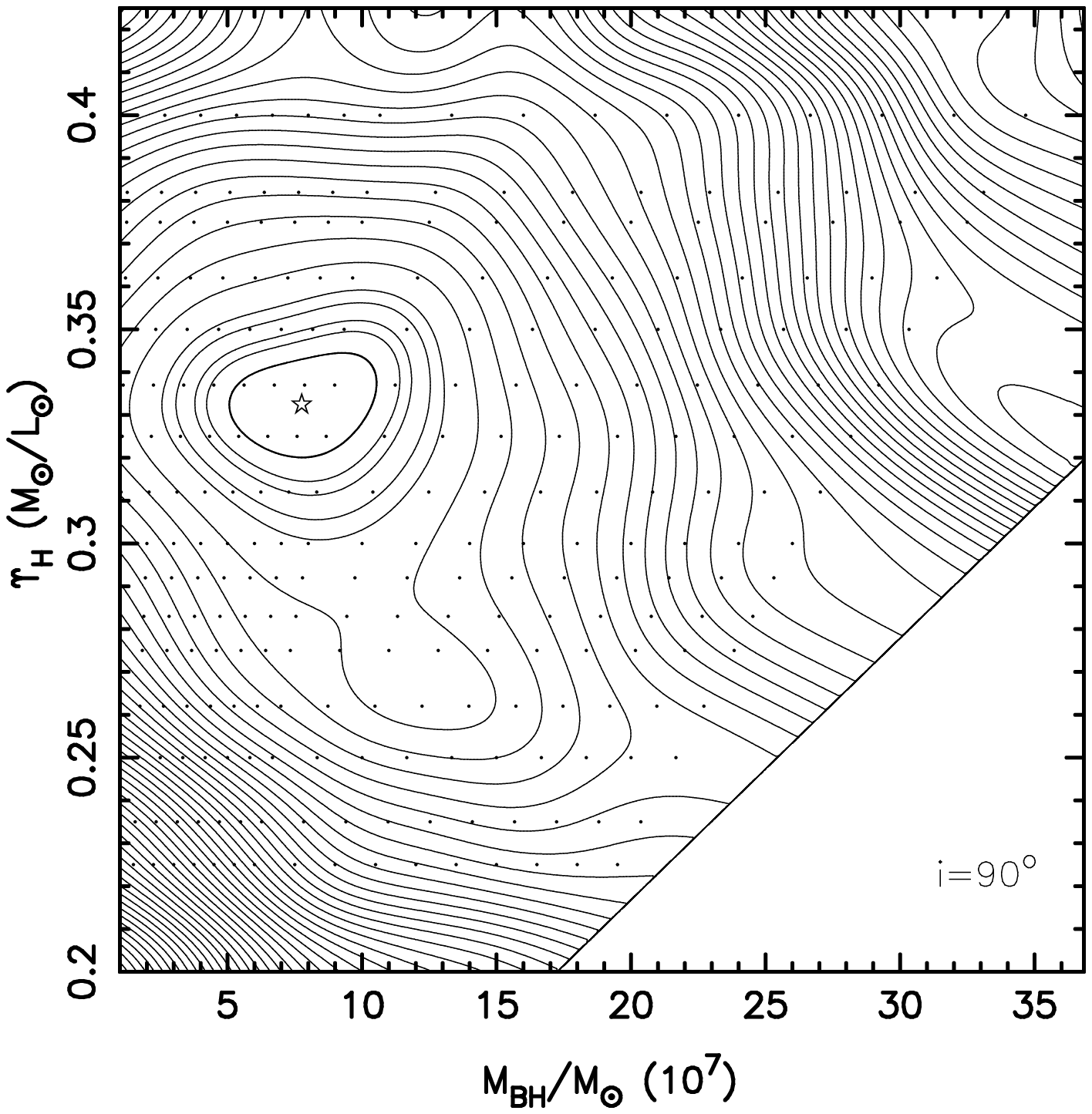}
\centering \includegraphics[trim=0.pt 0.pt 0.pt 0.pt, width=0.35\textwidth]{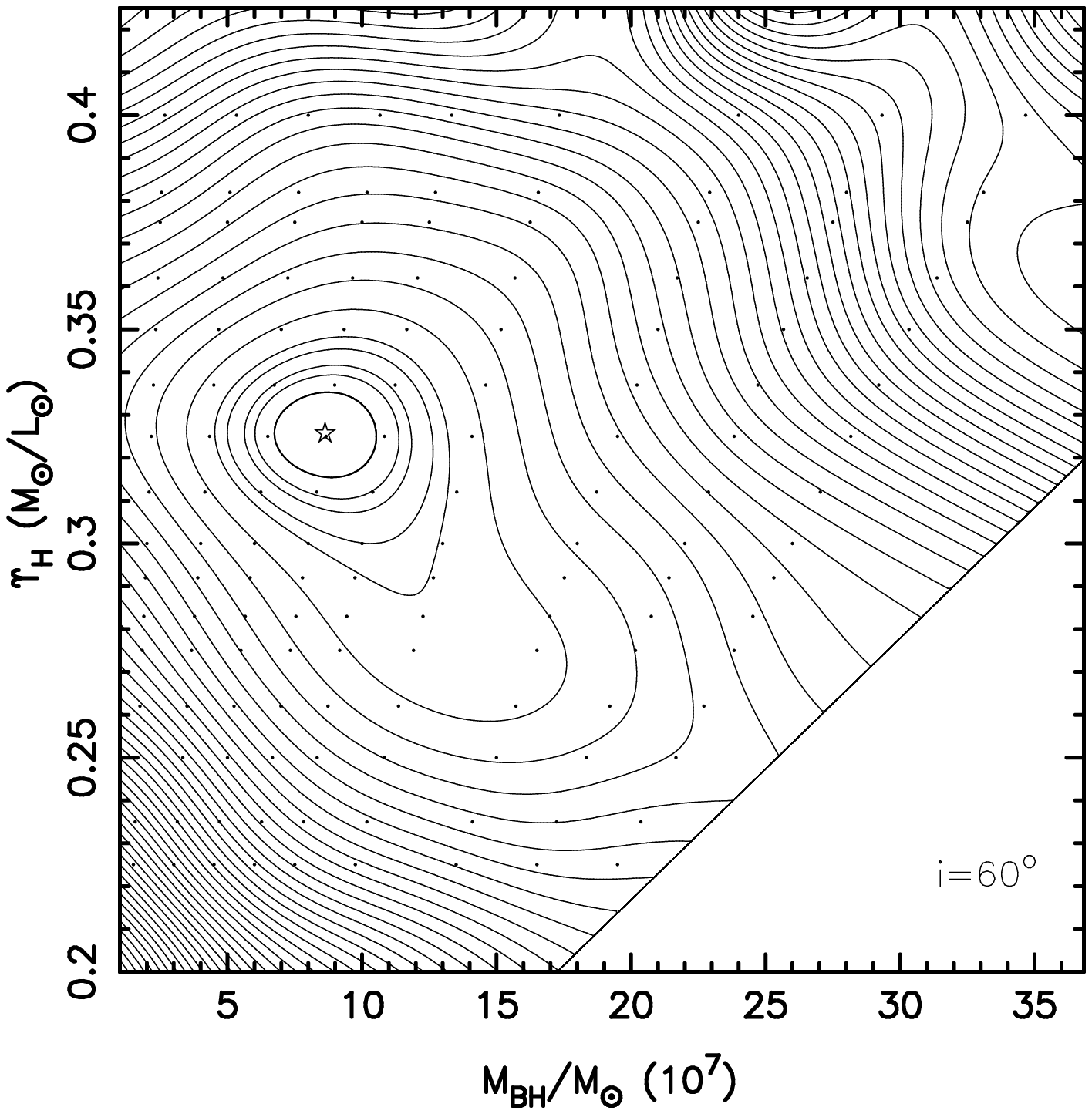}
\centering \includegraphics[trim=0.pt 0.pt 0.pt 0.pt, width=0.35\textwidth]{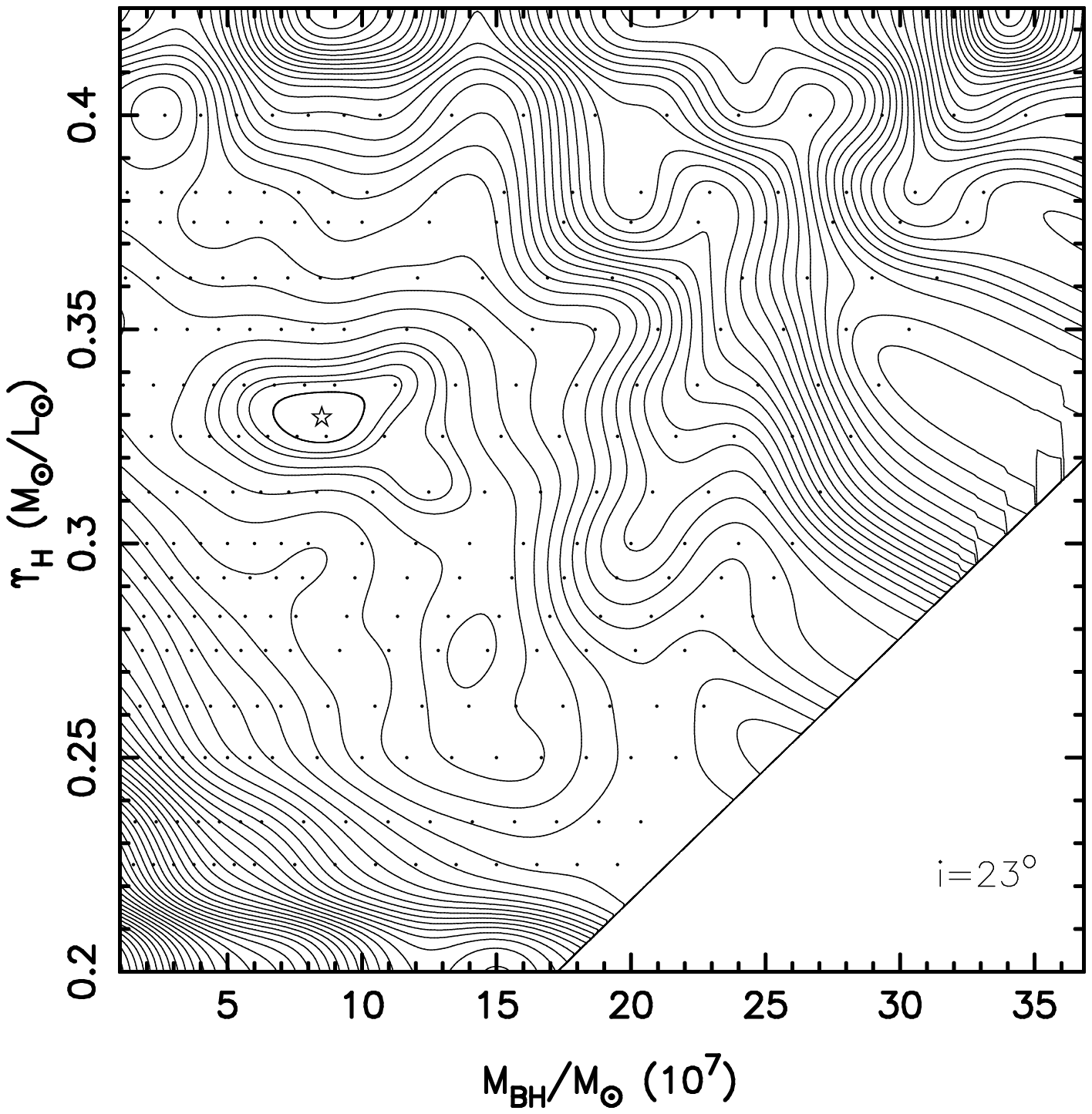}
\caption{Contours of constant $\chi^2_{\rm kin}$, where $\chi^2$ is
  computed only from the fit to the kinematic constraints for models
  with inclination to the line of sight $i=90\deg$. The stars mark the
  location of the best-fit values for \mbh and $\Upsilon_H$. The first
  6 contour levels in all three plots correspond to $1\sigma,
  2\sigma, \ldots, 6\sigma$ confidence intervals respectively, with
  subsequent contours being equally spaced.  }
\label{fig:2dChi2_kin}
\vspace{0.2cm}
\end{figure}

 
Figure~\ref{fig:2dChi2_kin} shows 2D contour plots of $\chi_{\rm kin}$
for the same sets of models as in Figure~\ref{fig:2dChi2}, with
inclination angles $i=90\deg, 60\deg, 23\deg$ (from top to bottom). We
emphasize that the model solutions that are used to generate
Figure~\ref{fig:2dChi2_kin} are identical to those used to generate
Figure~\ref{fig:2dChi2} -- only the quantities used in computing the
$\chi^2$ of the fit to the data differ.
 
\begin{figure}
\centering \includegraphics[trim=0.pt 0.pt 0.pt 0.pt, width=0.3\textwidth]{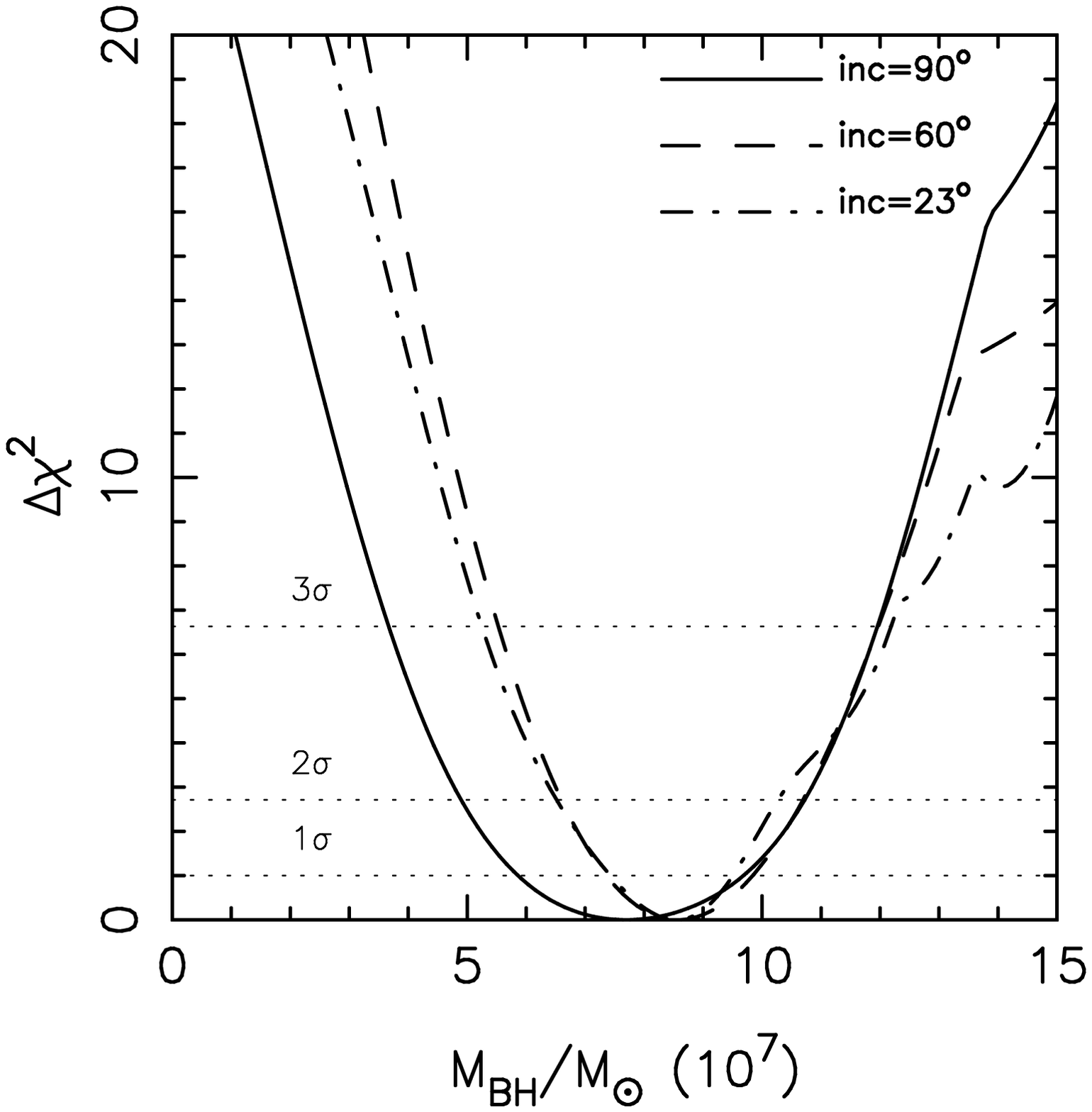}
\centering \includegraphics[trim=0.pt 0.pt 0.pt 0.pt, width=0.3\textwidth]{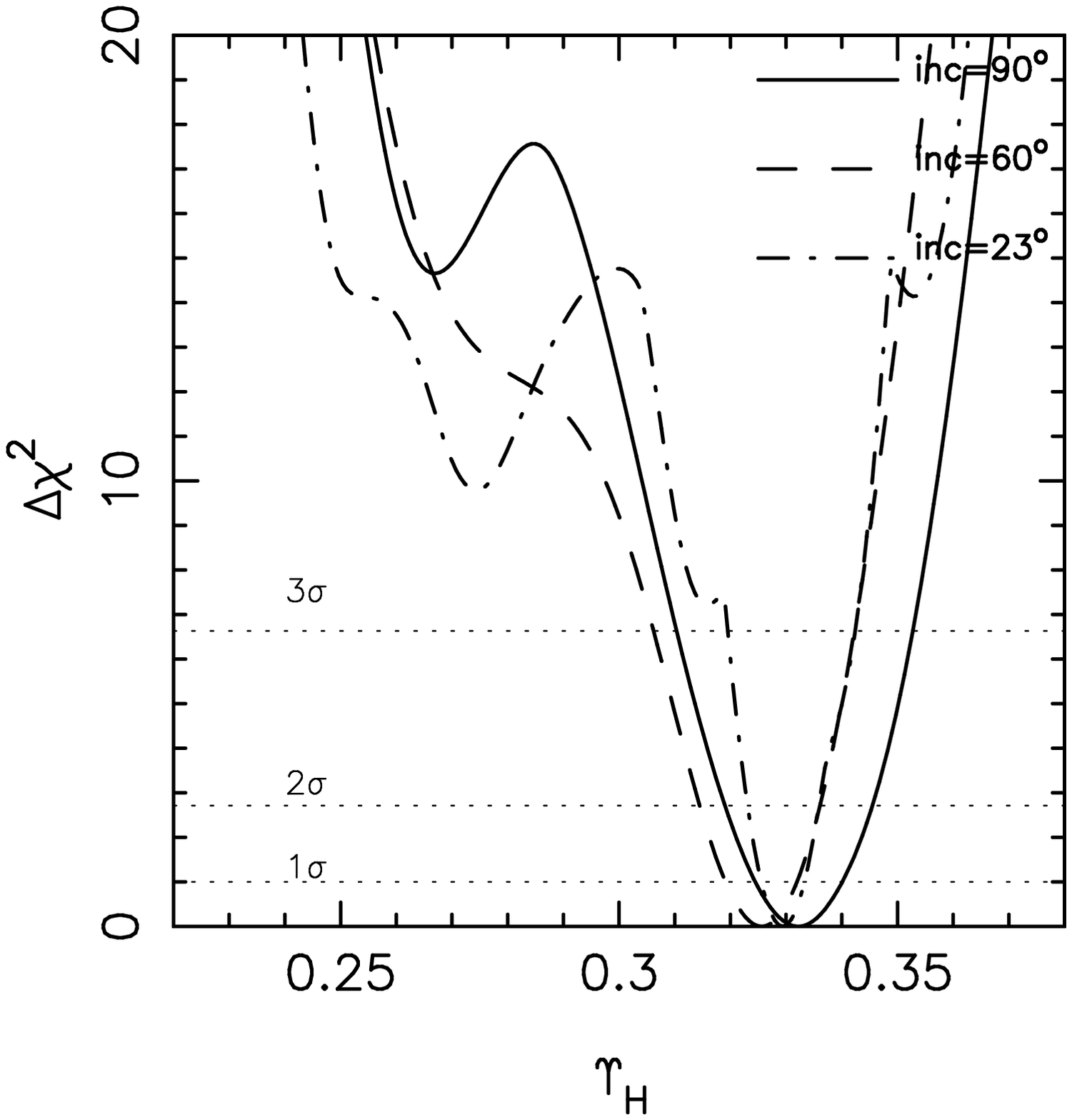}
\caption{Similar to Fig.~\ref{fig:1dChi2}, $\Delta \chi^2_{\rm kin}$
  curves marginalized over $\Upsilon_H$ (top) and marginalized over
  $\mbh$ (bottom) in Fig.~\ref{fig:2dChi2_kin}.}
\label{fig:1dChi2_kin}
\end{figure}

Figure~\ref{fig:1dChi2_kin} shows the same data marginalized over
$\Upsilon_H$ (top) and marginalized over $\mbh$ (bottom). It is clear
from these two figures that the best-fit solutions obtained using
$\chi^2_{\rm kin}$ give both a larger $\mbh \sim 8\times 10^7~\msun$
and a larger $\Upsilon_H \sim 0.33~\msun/\lsun$. We will discuss the
possible causes for this in Section~\ref{sec:robust} and
Section~\ref{sec:discuss}.

\subsection{Dependence on Inclination Angle of Bulge}
\label{sec:inclination}

\begin{figure}
\centering \includegraphics[trim=0.pt 0.pt 0.pt 0.pt, width=0.3\textwidth]{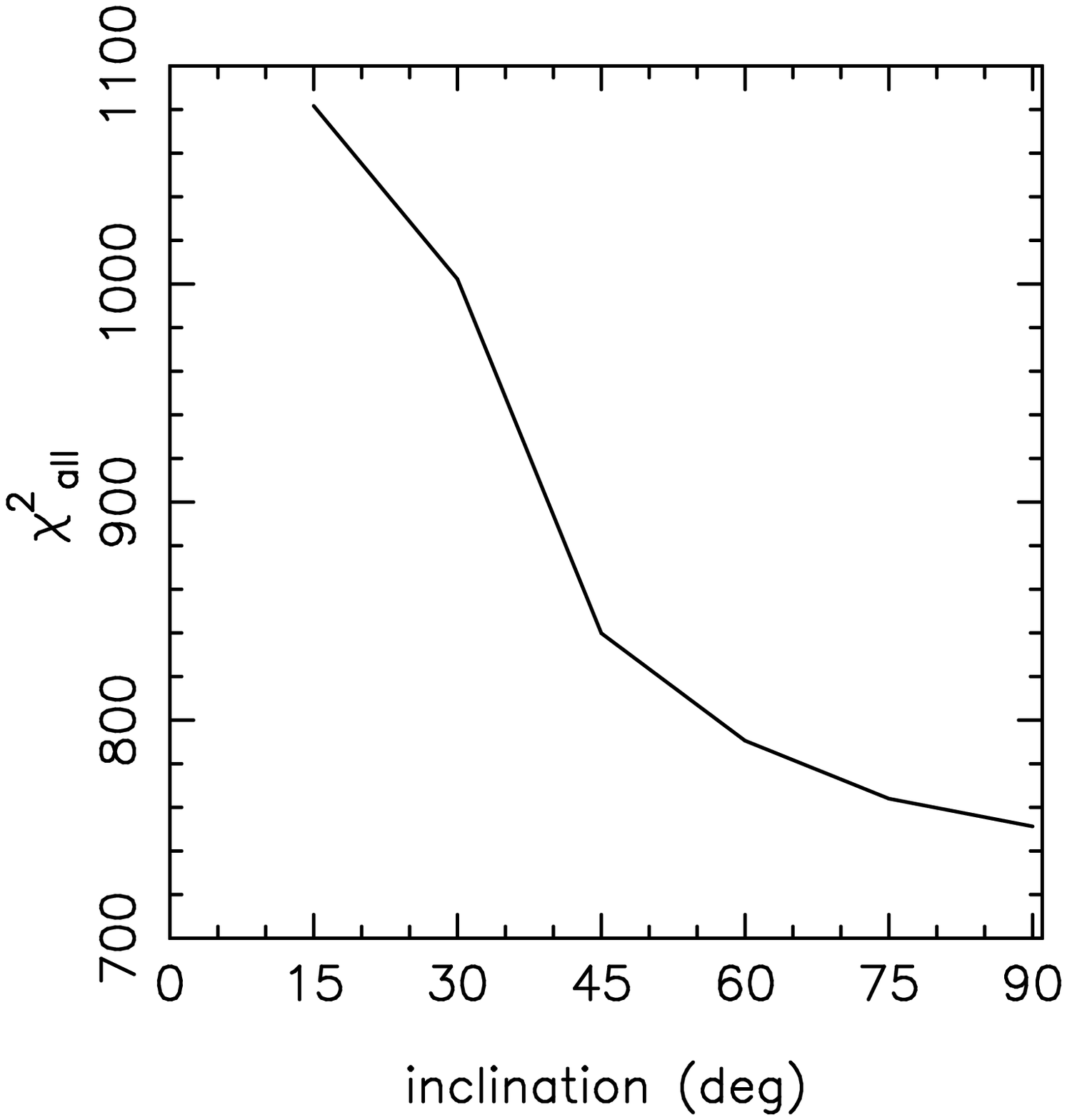}
\caption{$\chi^2_{\rm all}$ as a function of model inclination for
  models with $\mbh= 5.2\times 10^7~\msun$ and
  $\Upsilon_H=0.31~\msun/\lsun$. The edge-on model ($i=90\deg$) gives
  the smallest $\chi^2_{\rm all}$. }
\label{fig:chi_inc}
\end{figure}

In contrast to the results of Paper~I, where $i=90\deg$ models gave an
upper limit of $\mbh <4\times10^7 \msun$ and $i=23\deg$ models gave a
best fit $\mbh$ between $4-5\times10^7 \msun$,
Figures~\ref{fig:1dChi2} and \ref{fig:1dChi2_kin} show that best-fit
$\mbh$ is relatively insensitive to inclination (although the actual
best-fit value depends on how $\chi^2$ was
computed). Table~\ref{tab:bestfit} shows that the minimum values of
$\chi^2_{\rm all}$ and $\chi_{\rm kin}^{2}$ for $i=90\deg$ are lower
than the corresponding $\chi^2$ values for the other two inclination
angles. We now try to constrain the inclination angle of the bulge of
NGC 4151 by assuming a constant BH mass of $\mbh=
5\times10^7~\msun$. With this value of $\mbh$ we compute orbit
libraries for four additional inclination angles: $15\deg$, $30\deg$,
$50\deg$, and $75\deg$. Figure~\ref{fig:chi_inc} shows the values of
$\chi^2_{\rm all}$ for models with this value of $\mbh$ and
$\Upsilon_H=0.31~\msun/\lsun$ as a function of the inclination of the
bulge to the line-of-sight. It is clear that $\chi^2_{\rm all}$
decreases steadily with increasing inclination angle with a minimum at
$i=90\deg$. (A similar dependence on inclination angle is obtained for
$\mbh= 10^8~\msun$, although all the models give much worse fits as
assessed by the $\chiall$ values.)

The inclination angle of $i=23\deg$ that is inferred from the
\ion{H}{1} velocity field is strongly disfavored. Since the disk of
this galaxy is clearly seen to be close to face-on, the edge-on
orientation of the bulge preferred by our models is rather puzzling:
the surface brightness isophotes of the bulge are nearly circular,
hence the best-fit edge-on orientation implies that the bulge is
nearly spherical in shape; but in this case the rotation axis of the
bulge must be misaligned with the rotation axis of the disk by $\sim
70 \deg$.

Previous determinations of the orientations of the various components
of NGC~4151 have shown a remarkable diversity. While the circular
\ion{H}{1} isophotes of the large-scale disk suggest the nearly
face-on orientation with $i=23\deg$ \citep{simkin_75} -- which
motivated our testing of that value of inclination -- the components
associated with the AGN have been inferred to lie further from our
line of sight.  Models of the bi-conical narrow-line region (NLR) in
the UV and optical \citep{das05,shimono10,crenshaw10}, and in the
near-IR \citep{ruiz03,storchi-bergmann10,mueller-sanchez11} have
typically found inclinations of $40\deg-50\deg$. Closer to the AGN,
\citet{gallimore99} fit the \ion{H}{1} absorption using a pc-scale
nuclear disk inclined at $i=50\deg$.  On even smaller scales, an
inclination of $20\deg$ for the AGN accretion disk was obtained from
modeling the X-ray spectrum \citep{takahashi02}, while
\citet{storchi-bergmann10} suggest that the radio jet in NGC~4151 lies
very close to the plane of the sky and is interacting directly with
circumnuclear gas in the plane of the galaxy (i.e., $i >
60\deg$). This lack of alignment between the different spatial scales
could be the consequence of a merger in the galaxy's recent past. It
has been claimed that the periodic variability in the nuclear activity
of NGC~4151 is indicative of a BH binary, with a separation on the
order of $10^3$~AU \citep{bon12}. While distinguishing between a
single BH and a close binary is far beyond the capability of our
dataset, the remnants of a merger could also contribute to the
complicated dynamics we observe on the size scales of the bulge. In
Section~\ref{sec:robust}, we will show that the kinematics in the
inner 5\arcsec\ show evidence for a bar in this region, implying that
the internal dynamics of the bulge are likely not that of a spherical
or axisymmetric system.  A full understanding of why the kinematics of
our model prefers an edge-on configuration to the more inclined
configuration requires the system to be modeled by a bar dynamical
modeling code, which does not exist at the present time.

Since the result of this section show that models with edge-on
inclination ($i=90\deg$) are greatly preferred over more face-on
models, in the rest of this paper we confine our analysis to models
with $i=90\deg$.

\subsection{Tests of Robustness and Kinematic Fits}
\label{sec:robust}

\begin{table*}[ht]
\caption{Results of Robustness Tests}
\centering
\begin{tabular}{lllllcccccc}\hline 
\multicolumn{1}{l}{$\chi^2$ type }&
\multicolumn{1}{l}{$N_o$}&
\multicolumn{1}{l}{$N_c$}&
\multicolumn{1}{l}{$\chi^2_{\rm min}$}&
\multicolumn{1}{l}{$\chi^2_{\rm red}$}&
\multicolumn{3}{c} {$\mbh (10^7~\msun$)}& 
\multicolumn{3}{c} {$\Upsilon_H$ ($\msun/\lsun$)} \\
\multicolumn{1}{l}{ } &
\multicolumn{1}{l}{ } &
\multicolumn{1}{l}{ } &
\multicolumn{1}{l}{ } &
\multicolumn{1}{l}{ } &
\multicolumn{1}{c} {best-fit} &
\multicolumn{1}{c} {1$\sigma$} &
\multicolumn{1}{c} {3$\sigma$}& 
\multicolumn{1}{c} {best-fit} &
\multicolumn{1}{c} { $1\sigma$} &
\multicolumn{1}{c} {$3\sigma$}\\
(1) & (2) & (3) & (4) & (5) & (6) & (7) & (8) & (9) & (10) & (11)\\
\hline\\
$\chi^2_{\rm all}$& 15092 & 1240& 641.9 & 0.52 &   4.68   & $^{1.09}_{-0.99}$ & $^{+2.61}_{-2.23}$& 0.31 & $\pm0.003$ &$\pm0.01$\\
                           &            &          &          &          &             &                           &                              &            &                     &  \\
$\chi^2_{\rm kin}$&15092 & 768  &170.2 & 0.22  & 7.32     &$^{+1.18}_{-1.05}$& $^{+3.82}_{-2.65}$& 0.31& $\pm0.01$ &$\pm0.03$\\
                            &           &         &          &           &            &                              &                               &          &                     &                  \\                 
\hline
\multicolumn{11}{c}{Fit includes NIFS data only}\\
\hline\\             
$\chi^2_{\rm all}$& 9936  & 960  & 583.1& 0.61   &  3.82   &$^{+1.00}_{-1.11}$ & $^{+1.55}_{-2.02}$& 0.32 & $\pm0.01$ &$\pm0.02$\\
                            &           &          &         &          &            &                               &                              &            &                      &                  \\
$\chi^2_{\rm kin}$& 9936 & 680   & 101.7& 0.14 & 3.70    & $^{+1.25}_{-1.26}$ & $^{+3.30}_{-2.0}$& 0.35 & $\pm0.01$ &$\pm0.03$\\
\hline
\end{tabular}
\label{tab:robustbestfit}
\vspace{0.5cm}
\end{table*}

VME04 showed that the solution derived from the Schwarzschild method
was sensitive to two important factors: (a) the size of the orbit
libraries used in obtaining the best-fit BH mass; and (b) the spatial
resolution of the kinematic data and whether or not they resolved the
sphere-of-influence of the BH. They showed that biased solutions could
result when the size of the orbit libraries was too small for the
available data.  They also showed that when the kinematic data did not
resolve the sphere-of-influence of the BH, spurious results could
arise.  In the next two sections we perform two tests to assess the
robustness of our solutions. In particular, our goal is to assess
which of the two values of $\mbh$ obtained previously (5$\times
10^{7}~\msun$ or 8$\times 10^{7}~\msun$)\ is more robust.

\subsubsection{Dependence on Orbit Library Size}
\label{sec:robust_size}

In this section, we construct a set of 288 models with $i=90\deg$ but
with orbit libraries containing $N_o=15092$ orbits (as compared to
$N_o=9936$ used thus far) to fit the same set of 1240 constraints that
were fitted in Section~\ref{sec:define_chi2}. Large orbit libraries
are constructed for 16 BH masses between 10$^4~\msun$\ and
15$\times10^{7}~\msun$\ and for 18 values of $\Upsilon_{H}$.

\begin{figure}
\centering \includegraphics[trim=0.pt 0.pt 0.pt 0.pt, width=0.3\textwidth]{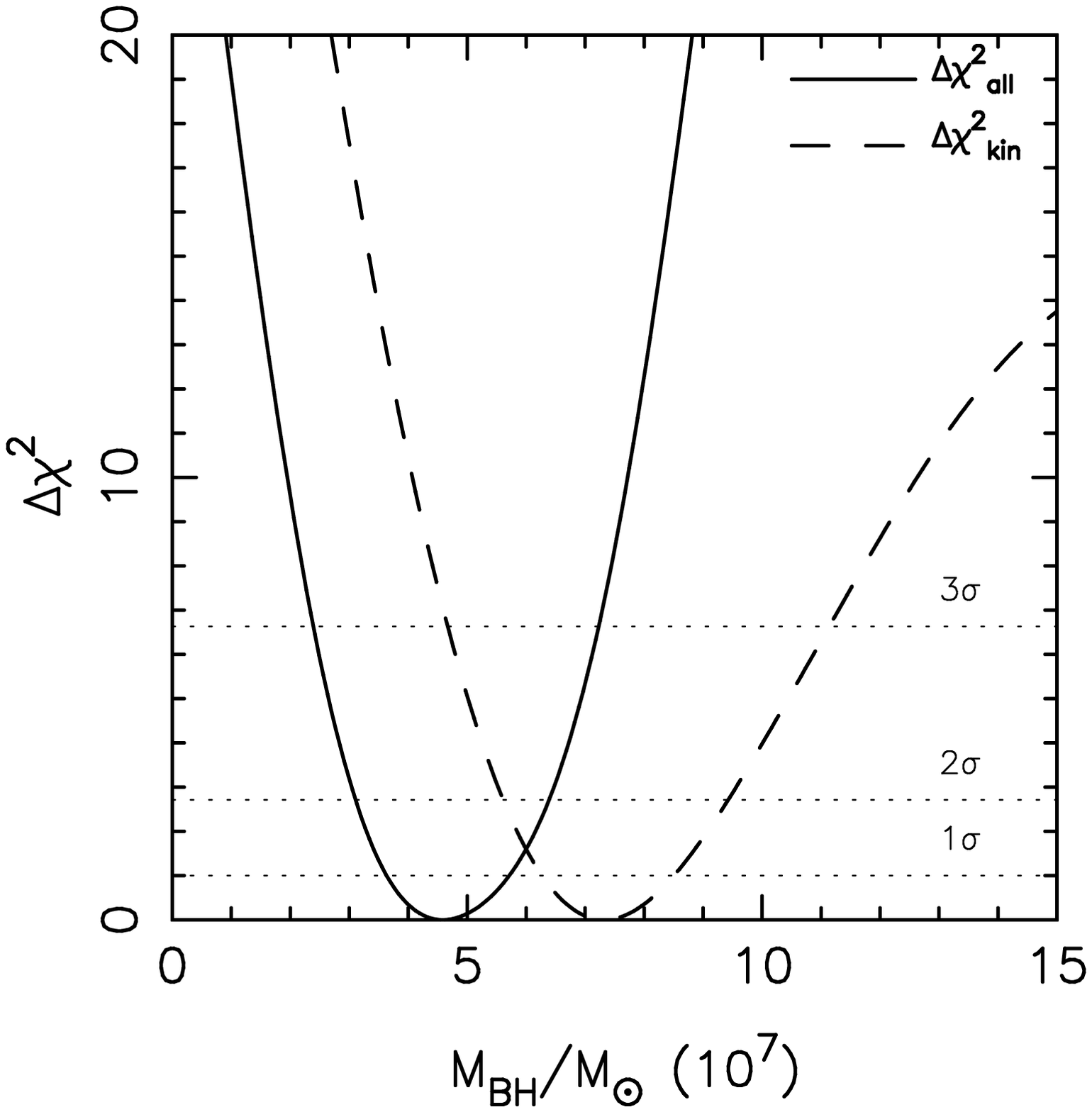}
\centering \includegraphics[trim=0.pt 0.pt 0.pt 0.pt, width=0.3\textwidth]{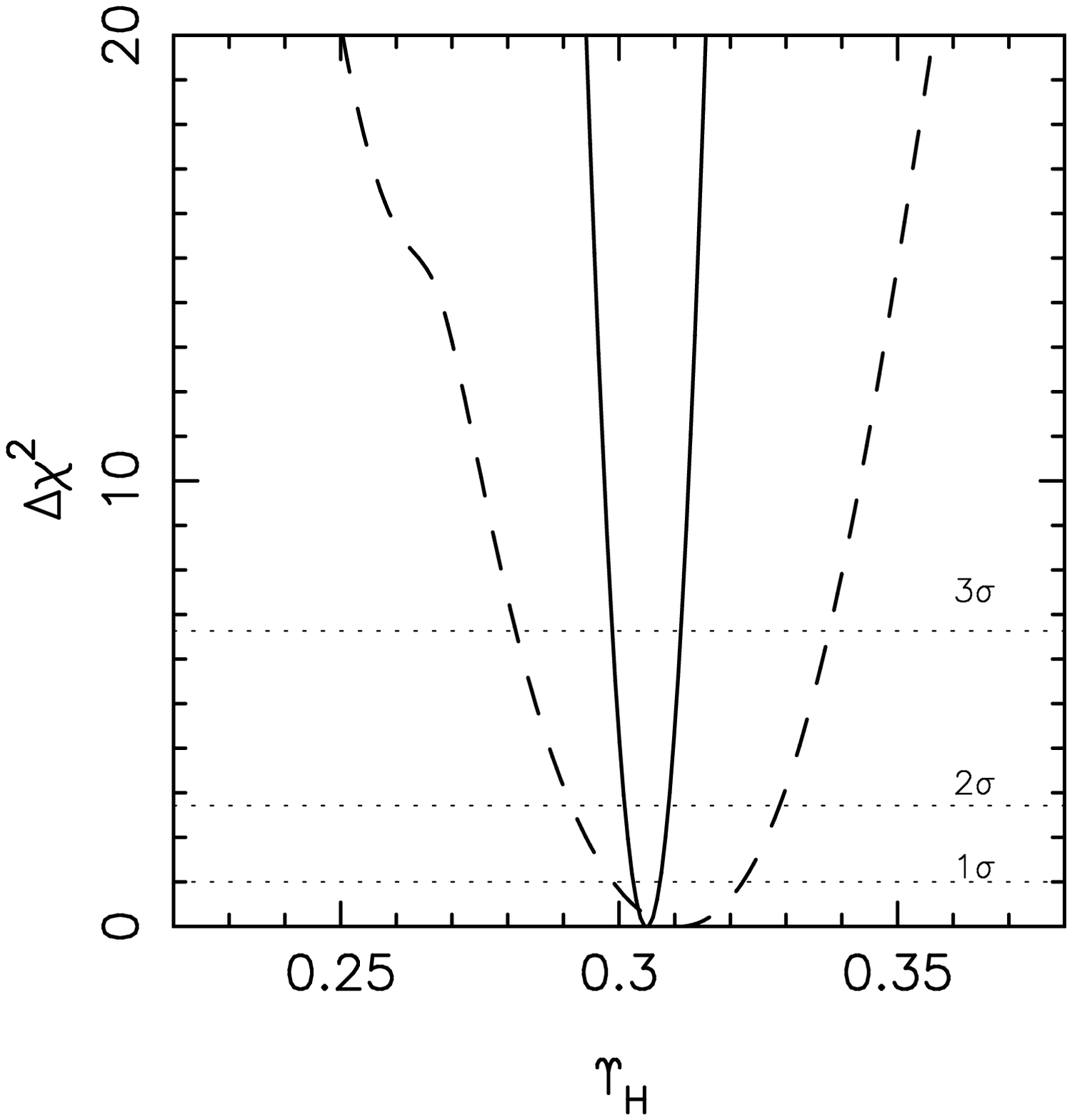}
\caption{1D $\Delta\chi^2$ obtained by marginalizing over $\Upsilon_H$
  (top) and marginalized over $\mbh$ (bottom) for models constructed
  using an orbit library of $N_o=15092$ orbits. In both panels
  $\Delta\chi^2_{\rm all}$ is shown by solid curves and
  $\Delta\chi^2_{\rm kin}$ is shown by dashed curves. The dotted
  horizontal lines indicate 1,2,3 $\sigma$ levels in $\Delta\chi^2$.}
\label{fig:1dChi_i9015k}
\vspace{0.5cm}
\end{figure}

Figure~\ref{fig:1dChi_i9015k} shows the 1D $\chi^2_{\rm all}$\ (solid
curves) and $\chi^2_{\rm kin}$\ (dashed curves) obtained by
marginalizing over the mass-to-light ratio $\Upsilon_H$ (top panel)
and obtained by marginalizing over $\mbh$ (bottom panel). The
resulting best-fit values for $\mbh$ and $\Upsilon_H$, and their
1$\sigma$ and $3\sigma$ errors, are shown in the top two rows of
Table~\ref{tab:robustbestfit}. The reduced $\chi^2$ values are given
in column five. It is clear from a comparison with results in
Table~\ref{tab:bestfit} that the best-fit $\mbh$ values are relatively
insensitive to an increase in the size of the orbit library by
$\sim$50\%. The best-fit value of $\Upsilon_H$ obtained by using
$\chi^2_{\rm kin}$ (dashed curves in the bottom panel of
Fig.~\ref{fig:1dChi_i9015k}) is somewhat lower than the value of
$\Upsilon_H= 0.332~\msun/\lsun$ obtained with the smaller library, but
is still within 2$\sigma$ of this value. The test with the larger
orbit libraries confirms the result we obtained with $N_o= 9936$
orbits per library, implying that our best-fit parameters are not
biased by having too small an orbit library. Furthermore, the
discrepancy between the best-fit values obtained using $\chi^2_{\rm
  all}$ and $\chi^2_{\rm kin}$ persists even with larger orbit
libraries.

\begin{figure*}
\centering \includegraphics[trim=0.pt 180.pt 0.pt 55.pt, clip, width=1.1\textwidth]{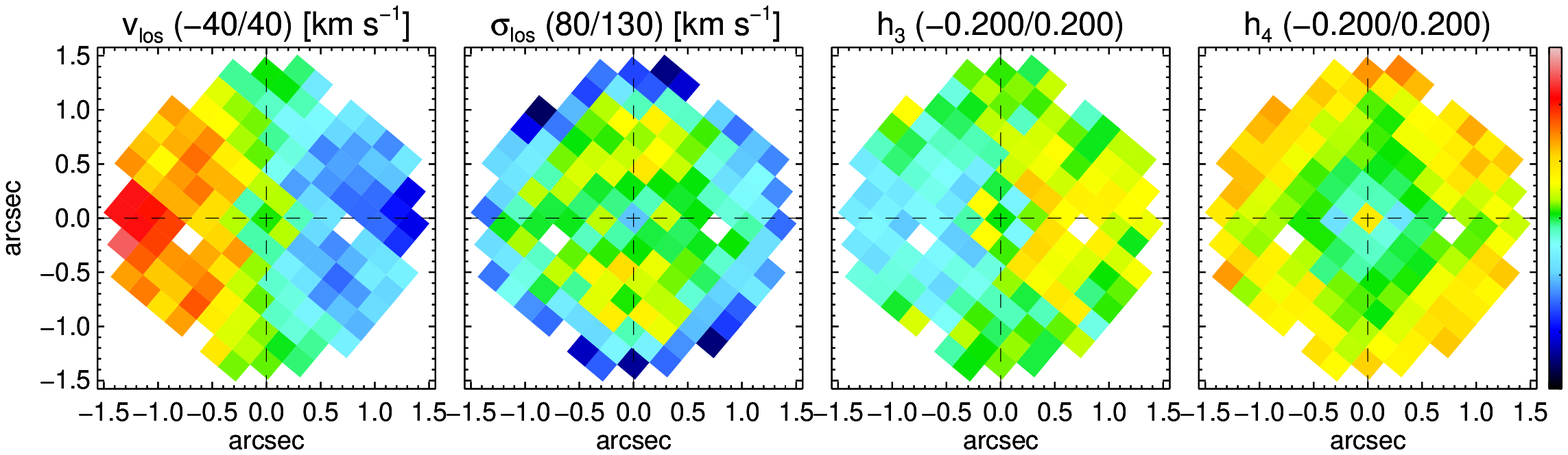}
\centering \includegraphics[trim=0.pt 180.pt 0.pt 55.pt, clip, width=1.1\textwidth]{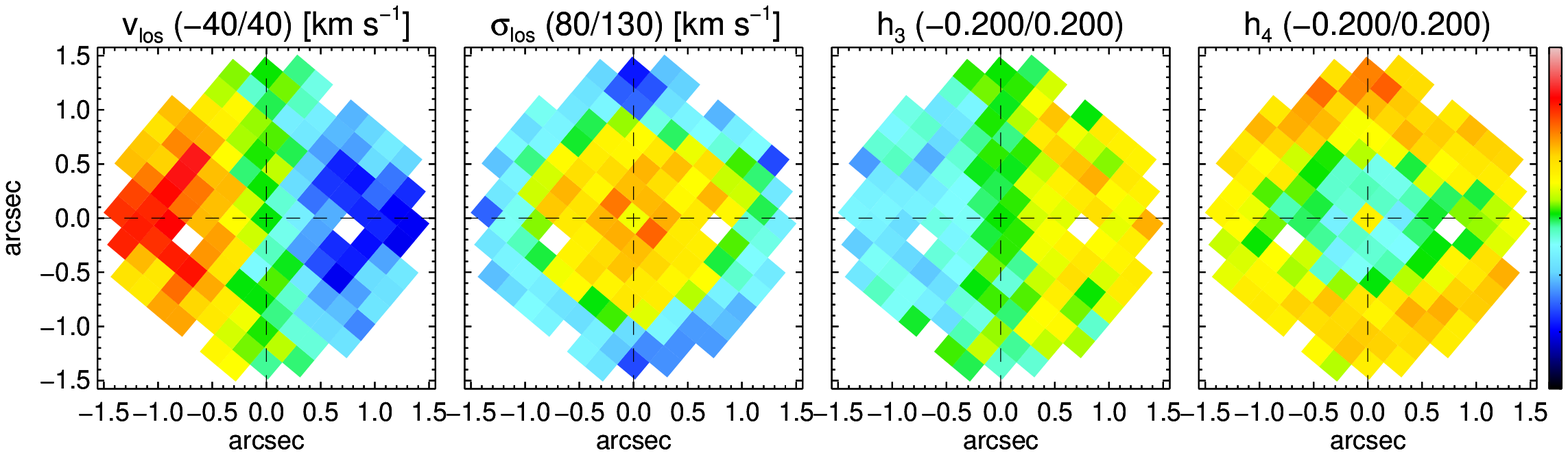}
\centering \includegraphics[trim=0.pt 180.pt 0.pt 55.pt, clip, width=1.1\textwidth]{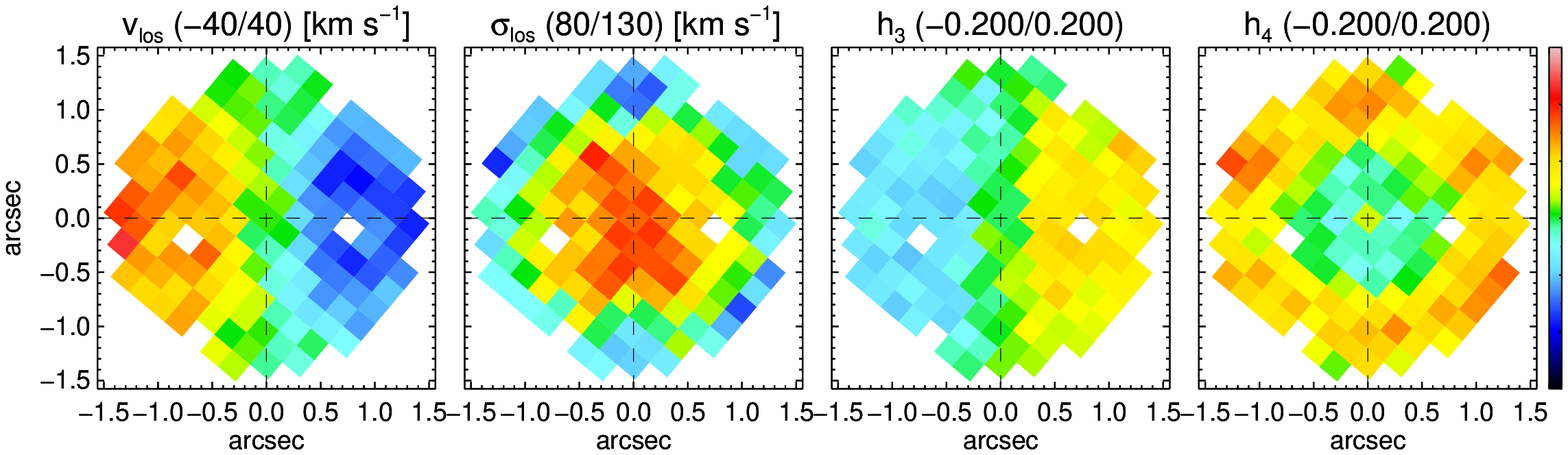}
\caption{Maps of fits to the NIFS line-of-sight velocity (\vlos),
  velocity dispersion (\siglos) and the 3rd and 4th Gauss-Hermite (GH)
  moments ($h_3, h_4$) for $\mbh=5\times10^7~\msun,
    \Upsilon_H=0.3~\msun/\lsun$ with $N_c=1240$, $N_o=15092$ orbits
    (top row) and for $\mbh=7.28\times10^7~\msun,
    \Upsilon_H=0.312~\msun/\lsun$ $N_c=1240$, $N_o=15092$ orbits
    (middle row). The bottom row shows the model velocity fields
    obtained when only NIFS constraints are included in the fit
    ($\mbh=4.16 \times10^7~\msun, \Upsilon_H=0.312~\msun/\lsun$,
    $N_c=960$, $N_o=9936$). These maps should be compared with the
  bi-symmetrized velocity fields in the bottom panels of
  Fig.~\ref{fig:vel_fields_orig}.}
\label{fig:kineNIFS_90}
\vspace{0.5cm}
\end{figure*}

Figure~\ref{fig:kineNIFS_90} shows examples of model fits to the 2
dimensional NIFS velocity fields (from left to right \vlos, \siglos,
and the GH moments $h_3$ and $h_4$) for the edge-on ($i=90\deg$) case
with $N_o=15092$. Note that while the best-fit solutions listed
  in Tables~\ref{tab:bestfit} \& \ref{tab:robustbestfit} are
  determined by marginalizing over one parameter in the 2D contour
  surface (which itself is smoothed with a kernel), the 2D grid of
  models is discrete. The derived best-fit value (shown as stars on
  the 2D $\chi^2$ maps of Figs.~\ref{fig:2dChi2} and
  \ref{fig:2dChi2_kin}) does not overlap with any model on the grid,
  but lies between 4 grid points. We therefore show velocity fields
  and 1D kinematics for the model whose parameters are closest to the
  best-fit solution.

In Figure~\ref{fig:kineNIFS_90}, the top row shows velocity fields
  for $\mbh=5\times10^7~\msun, \Upsilon_H=0.3~\msun/\lsun$ (the model
  closest to the best-fit obtained with \chiall\, for $N_o=15092$) and
  the middle row shows velocity fields for
  $\mbh=7.28\times10^7~\msun$, $\Upsilon_H=0.312~\msun/\lsun$ (the
  model closest to the best-fit obtained with \chikin\, for
  $N_o=15092$). (The bottom row of this figure is for a model with
  $\mbh=4.16\times10^7~\msun$, $\Upsilon_H=0.312~\msun/\lsun$,
  obtained when only NIFS kinematical data are fitted, and will be
  discussed in Section~\ref{sec:robust_nuclear}.)

The fitted velocity in Figure~\ref{fig:kineNIFS_90} should be compared
with the bi-symmetrized velocity fields in the bottom row of
Figure~\ref{fig:vel_fields_orig}. The four white pixels in each panel
of the figure were not fitted because it was determined that their
extremely high velocity dispersion values ($>130$~\kms) were spurious
(see Fig.~\ref{fig:vel_fields_orig}) since their values were larger
than the central velocity dispersion despite being far from the
center.

In Figure~\ref{fig:kineNIFS_90} the odd-moments of the LOSVD (\vlos,
$h_3$) are fairly well fitted by both the low $\mbh$ model (top row)
and the high $\mbh$ model (middle row). However, the NIFS velocity
dispersion \siglos\ for the model with $\mbh=5\times10^7~\msun$,
$\Upsilon_H=0.3~\msun/\lsun$ (top row) is much lower than \siglos\ in
the data (bottom of Fig.~\ref{fig:vel_fields_orig}). The model with
the larger value of $\mbh=7.28\times10^7~\msun$ does a slightly better
job of reproducing the higher \siglos\ values within $\pm 0\farcs5$,
but fails to fit the \siglos\ at the edges of the NIFS field, where
model values are significantly lower (i.e. bluer pixel colors) than
the observed values seen in Figure~\ref{fig:vel_fields_orig}
(bottom). This is because \siglos\ from NIFS in the inner 1\farcs5
region is overall larger than that obtained with KPNO ($\siglos
\sim75$~\kms) and MMT ($\siglos\sim90$~\kms).  Inconsistent velocity
dispersion values in the {\it same physical region} drives the
solution to fit the data with the smallest error bars (i.e., MMT data)
thereby under-estimating \siglos\ from NIFS. Also note that neither
model is able to fit both the low central and high outer $h_4$ values
seen in the bi-symmetrized NIFS velocity fields
(Fig.~\ref{fig:vel_fields_orig} bottom row, right-most panel).

\begin{figure*}
\centering \includegraphics[trim=0.pt 0.pt 0.pt 0.pt, width=0.7\textwidth]{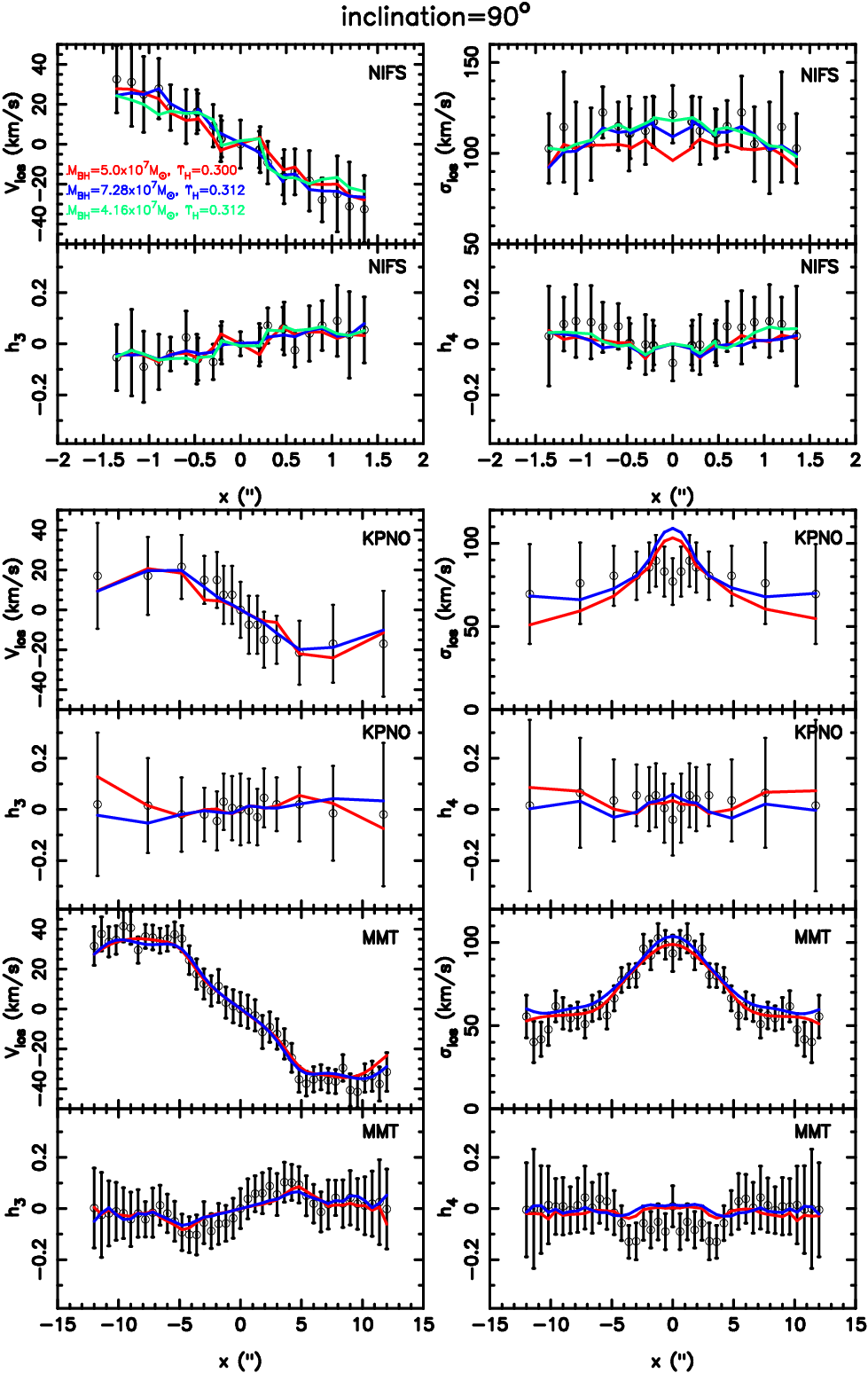}
\caption{Fit to observed kinematics \vlos, \siglos, $h_3$, $h_4$ for:
  NIFS apertures along the kinematic major axis (top four panels),
  KPNO slit (middle four panels), and MMT slit (bottom four panels),
  for inclination angle $i=90\deg$. Curves show fits for three different models with
  $\mbh$ and $\Upsilon_H$ values of: $5\times10^{7}~\msun$ and
  $0.30~\msun/\lsun$ (red); $7.28\times10^{7}~\msun$ and
  $0.312~\msun/\lsun$ (blue); $4.16\times10^{7}~\msun$ and
  $0.312~\msun/\lsun$ (green, fit to NIFS kinematics only).}
\label{fig:kinematics_90_1D}
\vspace{0.5cm}
\end{figure*}

The difference between the model fits and the data is more clearly
seen in Figure~\ref{fig:kinematics_90_1D} which shows the
one-dimensional line-of-sight velocity (\vlos), velocity dispersion
(\siglos) and the GH moments of the LOSVD ($h_3$ and $h_4$) that we
obtain for the edge-on ($i=90\deg$) model. The red and blue curves
show the best fits for same models as in the top and middle panels of
Figure~\ref{fig:kineNIFS_90} with values \mbh and $\Upsilon_H$
indicated by the line legends in the top left panel of the first row
(these two models used libraries with $N_o=15092$ orbits). The green
curves in the top four NIFS panels are obtained when only the NIFS
data are fitted (with $N_o=9936$), this model is discussed in the next
section.

The open circles in the panels show the three different kinematic
datasets used in the modeling, along with their 1$\sigma$
uncertainties. (The top four panels show kinematics in the NIFS
spaxels lying closest to the major axis of the model; the middle four
panels show data from the KPNO long-slit; the bottom four panels show data
from the MMT long-slit.)
 
The most striking feature of the fits is that the red curve for $\mbh
= 5\times 10^7~\msun, \Upsilon_H=0.3~\msun/\lsun$ (red curve) --- the
model closest to the best-fit solution derived from the minimum in
$\chi^2_{\rm all}$ (from Figure.~\ref{fig:1dChi2}) --- clearly {\it
  underestimates} the NIFS velocity dispersion (top right panel) for
the central-most points. The uniformly large central $\sigma_{\rm
    los}$ values in the inner $\pm1\arcsec$ region are much better
  fitted by larger values of $\mbh=7.28\times10^7\msun$ and
  $\Upsilon_H=0.312~\msun/\lsun$ (blue curve). Both the models do an
equally good job of fitting \vlos, \siglos, and $h_3$ from the MMT
data, but overestimate \siglos\ from the KPNO data.

Another important point to note is that the values of the GH
parameters $h_4$ are {\em negative} in the innermost regions of each
of the three datasets. Neither model is able to fit the negative $h_4$
values measured with the MMT spectrograph nor the central dips to
negative values in $h_4$ seen in the NIFS data and KPNO data.
 
In spherical and axisymmetric galaxies, $h_4$ is negative when the
velocity distribution is tangentially biased; it is positive when
orbits are predominantly radial; and it is zero when the velocity
distribution in isotropic \citep{vandermarel_franx_93,gerhard_93}.
However, negative $h_4$ values have also been associated specifically
with the presence of a bar. In recent $N$-body simulations of barred
galaxies by \citet{brown_etal_13}, the LOSVDs have negative $h_4$
values in the region where the bar dominates. The negative $h_4$
parameters arise due to the kinematic properties of the special orbits
that constitute the bar, and due to the fact that from an external
observer's point of view, the bar has a large tangential velocity
component due to its pattern speed. Even after a bar buckles and forms
a boxy bulge, negative $h_4$ values can be seen in face-on systems
\citep{debattista_etal_05}.

Another signature of the kinematics associated with a nuclear bar is
undulations in the rising part of the inner rotation curve
\cite[][]{bureau_athanassoula_05}, which we also see in the MMT data
for NGC~4151. Finally, while $h_3$ is always anti-correlated with the
line-of-sight velocity \vlos\ in axisymmetric systems,
\citet{bureau_athanassoula_05} show that $h_3$ is {\it correlated}
with \vlos\ along the major axis of the bar. For the data obtained
with KPNO (where the slit lies along the major axis of the large scale
bar) $h_3$ is correlated with \vlos\ over the radial range $\pm
2\arcsec$ (Figure~\ref{fig:kinematics_90_1D}), providing additional
kinematical evidence for a bar.

Therefore it appears that despite the very circular photometric
contours of the bulge in NGC~4151, and the visual appearance of a
rather weak bar at radii outside the bulge, there are several clear
kinematical signatures of a small scale bar in the vicinity of the
nuclear BH. This is consistent with the properties of ``barlens''
  features that have recently been identified in real and simulated
  galaxies \citep{laurikainen11,athanassoula14}. As pointed out by
  \citet{athanassoula14}, barlenses frequently masquerade as classical
  bulges because of their nearly circular central isophotes,
  especially when viewed nearly face-on, but have bar-like
  kinematics.

The combination of bars and BHs can have significant implications for
the central dynamics of galaxies. 
\citet{brown_etal_13} analyze a suite of $N$-body simulations of
barred galaxies in which point masses representing BHs were grown
adiabatically. They find that the growth of a BH of a given mass
causes a 5-10\% larger increase in $\sigma_{\rm los}$ in a barred
galaxy than in an otherwise identical axisymmetric
galaxy. \citet{hartmann_etal_13} find that if a disk galaxy with a
bulge has a pre-existing black hole, the formation and evolution of a
bar can result in a 15-40\% increase in the central velocity
dispersion. Both studies find that the increase in the observed
line-of-sight velocity dispersion in the barred systems is a
consequence of three separate factors: (a) mass inflow due to angular
momentum transport by the bar, (b) velocity anisotropy due to the
presence of bar orbits, and (c) weak dependence on orientation of
position angle of the bar.

If NGC~4151 does have a bar and is thus non-axisymmetric, that fact
also has implications for our modeling results. In stellar dynamical
modeling of axisymmetric galaxies, fitting a negative $h_4$ requires a
larger fraction of tangential orbits, which contribute little to the
line-of-sight velocity dispersion. Fitting a given velocity dispersion
with such an orbit population will imply a larger enclosed mass than
would be required if $h_4$ was zero or positive (which is typically
the case in axisymmetric systems with BHs). \citet{brown_etal_13}
argue that if an axisymmetric stellar dynamical modeling code is used
to derive the value of $\mbh$ in a barred galaxy with a high central
$\siglos$ and a negative $h_4$, the central BH mass will be
systematically overestimated. If the large values of $\siglos$ in the
nuclear regions and the negative $h_4$ values in the MMT data are a
consequence a nuclear bar, then the discrepancy between $\chikin$ and
$\chiall$ could reflect this predicted bias in $\mbh$.
  
\subsubsection{Restricting to Constraints from NIFS}
\label{sec:robust_nuclear}

In this section we examine the consequences of fitting only the NIFS
kinematical data (ignoring the kinematical constraints at larger radii
obtained from the lower spatial resolution long-slit spectra from KPNO
and MMT). There are three reasons why the NIFS data alone may provide
better constraints on the mass of the BH.

First, the NIFS data have higher spatial resolution (0\farcs2) than
the long slit data (1\arcsec-2\arcsec). For a $\mbh \sim 5\times
10^7~\msun$, the sphere-of-influence of the BH (assuming $\sigma_c =
116$~\kms; see Section~\ref{sec:msigma} below) is $\sim$16~pc, which
corresponds to $\sim$ 0\farcs24 for our assumed distance of 13.9~Mpc
for this galaxy. Therefore the NIFS data used in our modeling are
barely able to resolve the sphere-of-influence of the BH, but are
quite close to doing so. If $\mbh \sim 7.5\times 10^7~\msun$, the
sphere-of-influence of the BH would be 26~pc (0\farcs38) and the BH
sphere-of-influence would be resolved by the NIFS data. The long slit
data do not provide constraints on $\mbh$ and are included mainly to
provide constraints on the mass-to-light ratio of the stars, which is
assumed to be independent of radius.

Second, as discussed previously, the NIFS velocity dispersion within
$\pm2\arcsec$ is significantly larger than the velocity dispersion on
same physical scale obtained from the long-slit data.  However, since
the error bars on the long-slit MMT data are about half those from the
NIFS data, the optimization code tries to
fit them better than the NIFS data. 

In Figure~\ref{fig:kineNIFS_90}
(top two panels), the need to fit the lower velocity dispersion values
from MMT manifests as low velocity dispersion values (blue) at the
edges of the NIFS field.  These low dispersion values outside the
sphere-of-influence of the BH also force the model to adopt a lower
mass-to-light ratio, which is then compensated for in the inner region
by requiring a much larger \mbh to fit the NIFS velocity
dispersion. Neglecting the long-slit data would allow the optimization
code to raise the mass-to-light ratio in the inner $\pm$2\arcsec\ 
region (where the velocity dispersion is quite flat).

Finally, in Figure~\ref{fig:kinematics_90_1D}, we see kinematic
evidence for the bar in MMT and KPNO long-slit data (negative $h_4$
values and undulations in the rising part of the rotation curve seen
in MMT data; correlation between $h_3$ and $\vlos$\ in KPNO data)
occurring on scales of 3\arcsec-5\arcsec, well outside the nuclear
region. Except for the central value of $h_4$ in the NIFS dataset
(which, as we see from Fig.~\ref{fig.spec}, is contaminated by the
AGN), all the other $h_4$ values are positive, which is what is
expected for orbits around a central BH. Neglecting the long-slit
kinematical data could help overcome biases introduced by the bar
kinematics.  Assuming that the BH mass should primarily be constrained
by kinematic constraints at the smallest radii we now test the effect
of not including the long-slit data in the NNLS optimization
problem. The expectation is that neglecting the long-slit data should
bring the best-fit solutions obtained from $\chi^2_{\rm all}$ and
$\chi^2_{\rm kin}$ closer to each other.

\begin{figure}
\centering \includegraphics[trim=0.pt 0.pt 0.pt 0.pt, width=0.3\textwidth]{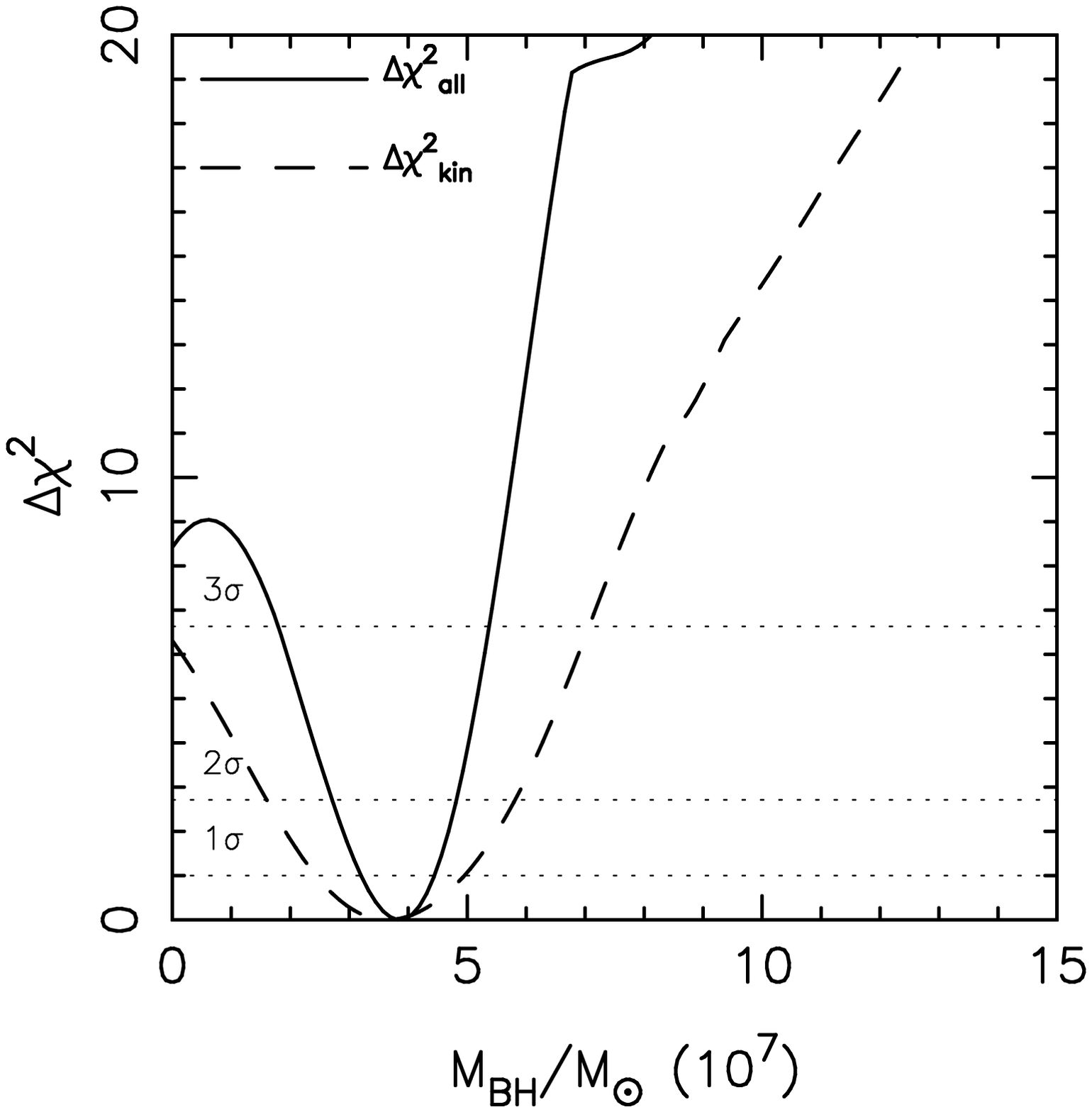}
\centering \includegraphics[trim=0.pt 0.pt 0.pt 0.pt, width=0.3\textwidth]{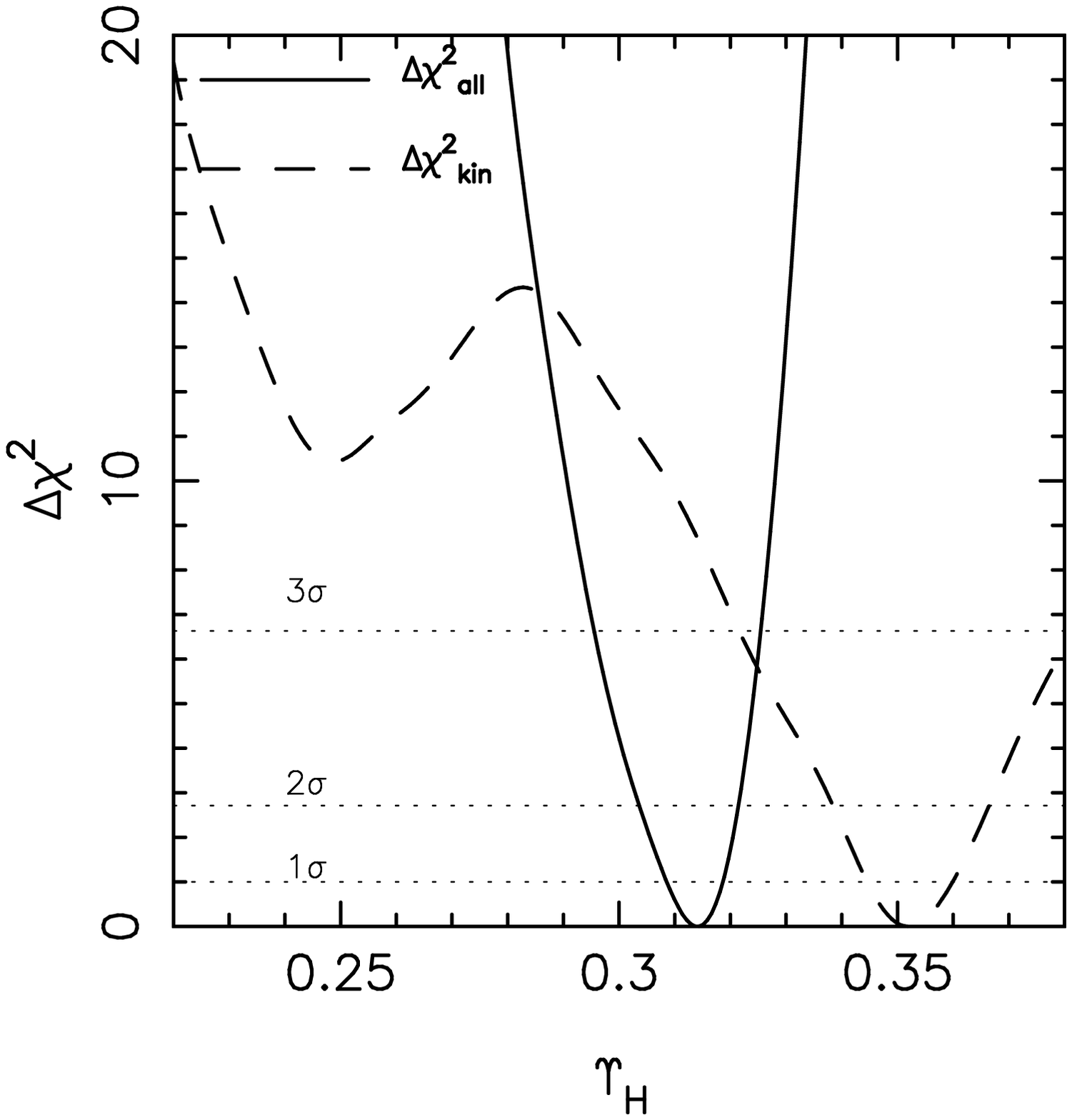}
\centering \includegraphics[trim=0.pt 0.pt 0.pt 0.pt, width=0.29\textwidth]{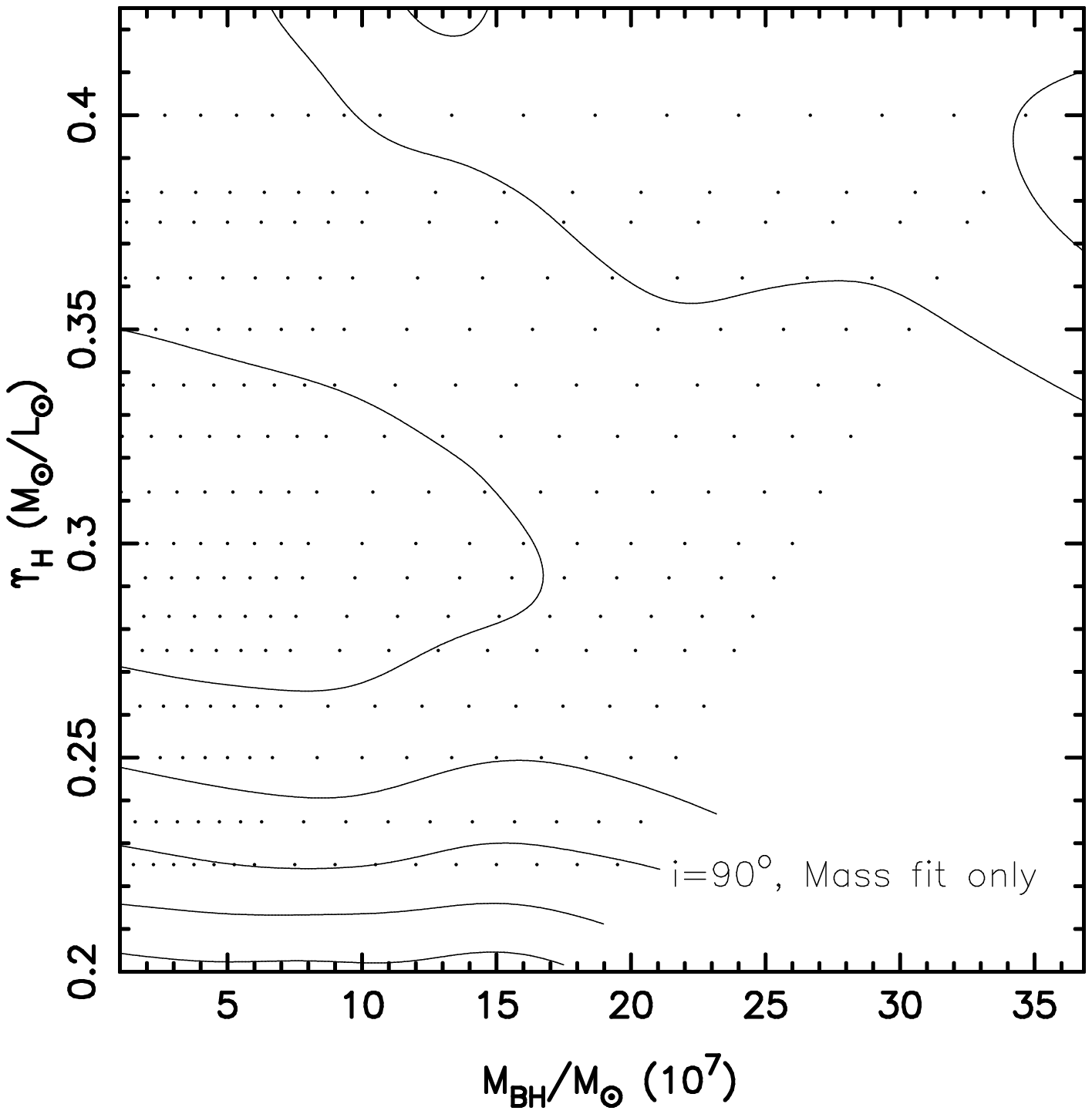}
\caption{Results of fitting NIFS kinematics only. 1D $\Delta\chi^2$
  obtained by marginalizing over $\Upsilon_H$ (top) and marginalizing
  over $\mbh$ (middle) for models constructed using an orbit library
  of $N_o=9936$ orbits.  $\Delta\chiall$ is shown by solid curves and
  $\Delta\chikin$ is shown by dashed curves. The bottom-most panel
  shows errors on the fit to the mass distribution (contours are
  spaced at 0.1\% error per constraint). Mass fit errors are uniform
  over a larger range of parameters and hence contribute negligibly to
  $\chiall$.}
\label{fig:nifs_only}
\vspace{0.5cm}
\end{figure}

We construct a full sets of models (18 values of $\mbh$ and 18 values
of $\Upsilon_H$) in which we only fitted the nuclear kinematical data
obtained with the NIFS spectrograph, with an orbit library of
$N_o=9936$.  Figure~\ref{fig:nifs_only} shows $\Delta \chi^2$ values
obtained by marginalizing over $\Upsilon_H$ (top) and by marginalizing
over $\mbh$ (middle). The solid curves show $\Delta \chi^2_{\rm all}$
while the dashed curves show $\Delta \chi^2_{\rm kin}$.  The dotted
horizontal lines show the 1$\sigma$, 2$\sigma$ and 3$\sigma$
confidence levels, respectively (for 1 degree-of-freedom). The top
plot shows that the minimum $\chi^2_{\rm kin}$ (dashed curve) has
decreased quite significantly and is now in agreement with the minimum
$\chi^2_{\rm all}$ (solid curve).  The best-fit values of $\Upsilon_H$
(middle panel) are still inconsistent within 1$\sigma$ but are
consistent within 3$\sigma$. The bottom-most panel shows the error on
the fit to the mass and surface brightness distribution. Contours are
spaced at intervals of 0.1\% error per constraint. It is clear that
the fit to the self-consistency constraints is uniform over a much
larger range of $\mbh$ and $\Upsilon_H$ values implying that the fit
to the mass is no longer a significant factor in determining the
location of the minimum.

The best fit black hole
mass and M/L ratio obtained using $\Delta \chi^2_{\rm all}$ and
$\Delta \chi^2_{\rm kin}$, along with the reduced $\Delta \chi^2$ and
1$\sigma$ and 3$\sigma$ error bars, are shown in the bottom two rows
of Table~\ref{tab:robustbestfit}. Since the best fit $\mbh$ values and
their 1$\sigma$ error bars from both methods of computing $\chi^2$ are
nearly the same we use the average of the two best-fit values as the
solution and sum their 1$\sigma$ errors in quadrature to obtain $\mbh
= 3.76 \pm 1.15$.  Figure~\ref{fig:nifs_only} shows that the
$\Upsilon_H$ values obtained from $\chi^2_{\rm kin}$ are consistent
with the best fit value obtained with $\chi^2_{\rm all}$ at the
3$\sigma$ level, but both are well within the expected range
$0.4\pm0.2~\msun/\lsun$. Therefore we use the average of the two
best-fit M/L values and sum their 3$\sigma$ errors in quadrature to
get $\Upsilon_H = 0.34 \pm 0.03$.

The bottom panel of Figure~\ref{fig:kineNIFS_90} shows a 2D
  kinematic map for a model with $\mbh=4.16\times10^7~\msun$ and
  $\Upsilon=0.312~\msun/\lsun$, with parameters closest to the
  best-fit solution obtained with $\chiall$ (see 3rd row of
  Table~\ref{tab:robustbestfit}).
  This model achieves adequately large values of $\siglos$ (see
  red/orange regions) over the central NIFS field by raising the M/L
  ratio very slightly from $\Upsilon=0.3$ to $\Upsilon=0.312$.  The
  1D kinematic fit of this model to the NIFS data along the kinematic
  major axis of the model is shown by the green curves in
  Figure~\ref{fig:kinematics_90_1D}, from which it is also clear that
  the lower value of $\mbh$ is able to fit the high velocity
  dispersion values even better than the high BH mass when the model
  attempted to fit both the NIFS and long-slit kinematic data.

This test largely confirms our hypothesis that the two main causes of
the large value of $\mbh$ obtained in the previous section were (a) inconsistencies
between $\siglos$ values from NIFS and the long-slit data in the same
spatial region, (b) the need to fit the low $h_4$ values and high
central $\siglos$\ values with a low M/L ratio.  Thus, we see that by
not including the kinematics data with poorer spatial resolution
(which are clearly unaffected by the BH as can be seen in
Figure~\ref{fig:kineNIFS_90}) we are able to simultaneously fit the
NIFS kinematics and the mass and surface brightness distributions with
a lower value of $\mbh$.

\subsection{The $M_{\rm BH}-\sigma$ Relation}
\label{sec:msigma}

In this section, we determine how the new NIFS kinematic measurements
in the nuclear region of NGC~4151 affect the galaxy's location in the
$M_{\rm BH}-\sigma$ relation. The aperture used to define the
appropriate $\sigma$ value for the relation is a historically
contentious issue \citep[e.g.][]{merritt_ferrarese_01,Tremaine02}, but
the NIFS data naturally lend themselves to one of those standards:
$\sigma_c$, which is the velocity dispersion within an aperture of
radius $R_e$/8, where $R_e$ is the bulge effective radius
\citep{ferrarese_merritt_00}. (Despite having to adopt one particular
standard, we note that \citet{merritt_ferrarese_01} found no
systematic differences between estimates of $\sigma_c$ (inside of
$R_e$/8) and estimates of $\sigma$ that extended out to $R_e$, likely
because of the steep radial surface brightness gradient in most
bulges.) The $R_e$ value for NGC~4151 has been measured to be $\approx
10\arcsec$ \citep{dong_robertis_06,bentz_etal_09,weinzirl_etal_09},
while the NIFS field-of-view extends to a radius of between $1\farcs5$
and $2\farcs1$. Thus, we sum the NIFS datacube across all of the
spaxels into a single spectrum, and fit for the dispersion using pPXF.

From the NIFS data, we find $\sigma_c = 116\pm3$~\kms. This is
somewhat larger than either the $89\pm13$~\kms\ measured by
\citet{ferrarese_etal_01}, or the $97\pm3$~\kms\ measured by
\citet{nelson_etal_04}. Both of those earlier measurements used the
Calcium triplet absorption lines, but whereas the former covered a
similar region of the galaxy as our NIFS spectra
($2\arcsec\times4\arcsec$), the latter used an aperture of
$1\arcsec\times6\farcs5$.  Previous studies of bulge kinematics have
found near-IR and optical data to give consistent results
\cite[see][and references therein]{kang_etal_13}, while the
discrepancies seen in some luminous IR galaxies are in the sense of
{\it smaller} $\sigma$ values from the CO-bandheads
\citep{rothberg_etal_13}.

Our new measurement of $\sigma_c=116\pm3$~\kms\ is $\sim$25\% above
the previous velocity dispersion measurements, while our best-fit
stellar dynamical estimate of $\sim3.76\pm1.15\times 10^7~\msun$ is
17\% below our earlier stellar dynamical estimate
\citep{onken_etal_07}. Previously, NGC~4151 had been an outlier from
the AGN $M_{\rm BH}-\sigma$ relation in the direction of low-$\sigma$
or high-$\mbh$.  The new $\sigma_c$ and $\mbh$ measurements helps to
bring NGC~4151 closer to the best-fit relation of \citet{woo_etal_13},
though it does remain on the low-$\sigma$ side.

Finally, we note that if the BH masses in barred galaxies have been
overestimated due to the use of axisymmetric dynamical modeling codes,
such galaxies would lie even further below the standard \msigma
relation than has been found previously \cite[][and references
therein]{graham_etal_11}. This demonstrates the importance of
developing robust bar dynamical modeling approaches in the future.

\section{Summary and Discussion}
\label{sec:discuss}

We have conducted AO-assisted near-IR integral field spectroscopy of
the local Seyfert galaxy NGC~4151, using the NIFS instrument on Gemini
North.  We used an axisymmetric orbit-superposition code to fit the
observed surface brightness distribution within 50\arcsec, NIFS
kinematics within $\pm$1\farcs5, and long slit kinematics along two
different position angles in order to estimate the best-fit values of
the mass of the BH ($\mbh$) and the mass-to-light ratio of the stars
($\Upsilon_H$). Models were constructed for 10-18 values of $\mbh$, 18
values of $\Upsilon_H$ and 3 different inclination angles.  The main
results of this paper are summarized below.

We use the $\chi^2$ estimator to determine the model which gives the
best-fit to the data but find that the solution depends on whether the
$\chi^2$ includes both self-consistency constraints (which are not
strictly speaking ``data'') and kinematic constraints or only
kinematic constraints. When we use both types of constraints and all
the available kinematical data (from low-resolution long-slit data and
high resolution NIFS data) the best-fit solution is $\mbh =
4.68\times 10^7~\msun$ and $\Upsilon_H \sim 0.30\msun/\lsun$, for a
model with $i=90\deg$. However, this best fit model gives a central
$\siglos$ that is too low to fit the data obtained with the NIFS
instrument. When we use $\chikin$ (determined by only considering the
fit to the kinematic data), the best fit model has $\mbh = 7.32\times
10^7~\msun$ and $\Upsilon_H\sim 0.31~\msun/\lsun$, and provides a
better fit to the nuclear kinematical data from NIFS spectrograph,
although it gives a slightly worse fit to the mass and surface
brightness distributions.

An interesting point worth noting is that \citet{hicks_malkan_08}
found a best-fit value of $\mbh=3^{+0.75}_{-2.2}\times 10^7$ in NGC
4151 when they fitted the kinematics of the H$_2$ line-emitting gas
within 1\arcsec\ of the center (see Figure 43 in their
paper). However, if they included the kinematics of gas within
2\arcsec\ they obtained a larger value of $\mbh \sim 8\times
10^7~\msun$, although with a somewhat worse reduced-$\chi^2$. It is
intriguing that both our stellar dynamical modeling and their gas
dynamical modeling, while preferring lower values of $\mbh \sim
3-5\times 10^7$, also suggest that {\em higher} values of $\mbh$ may be
obtained when lower spatial resolution data are included in the fit.
  
Models are generated for three different inclination angles of the
bulge $i=23\deg, 60\deg, 90\deg$ and show that the best-fit $\mbh$ and
$\Upsilon_H$ values are relatively insensitive to inclination
angle. We examine an additional 4 inclination angles for fixed BH mass
values ($\mbh = 5\times 10^7, 10^8$) and fixed mass-to-light ratio
($\Upsilon_H=0.3$). We find that edge-on models ($i=90\deg$) give the
smallest $\chi^2$ values, while models with the inclination of the
large scale disk ($i=23\deg$) are strongly disfavored. This suggests
that the bulge must be nearly spherical but its rotation axis is
significantly misaligned from the rotation axis of the disk. However,
it is well known from previous work \citep{krajnovic_etal_05,
  lablanche_etal_12} that inclination of a spheroid is extremely
difficult to determine via dynamical modelling, especially in the
presence of a bar. This determination of the inclination of the bulge
should therefore be treated with caution.

A detailed examination of the kinematics in the inner 5\arcsec\ of
NGC~4151 shows evidence for bar kinematics, which manifest as high
values of $\siglos$, low or negative values of $h_4$ and $h_3$ values
which are correlated with $\vlos$. We hypothesize that the discrepancy
between the $\chi^2$ obtained from including all the constraints and
that obtained by only including the kinematic constraints is likely a
consequence of two factors: (a) a stellar bar, which is not possible
to model with our axisymmetric code, and (b) inconsistencies between
the velocity dispersion values obtained from the low-spatial
resolution long-slit data and the high spatial resolution NIFS data
over the same spatial range. This hypothesis was tested by fitting
only the NIFS data.

When we fit only the kinematical data obtained with the NIFS
spectrograph (neglecting all kinematical constraints beyond $\pm
1\farcs5$ - i.e those data that show kinematical evidence for a bar
and data that is discrepant with high-resolution NIFS data), the
non-negative optimization problem was able to find a much better fit
to the NIFS data and the same best-fit $\mbh$ was obtained by both
methods of computing $\chi^2$.
  
The stellar-dynamical modeling carried out in this paper gives a
best-fit value of $\mbh = 3.76\pm1.15\times10^7\msun\,$ (obtained by
averaging over the $\chiall$ and $\chikin$ values in the lower two
rows of Table~\ref{tab:robustbestfit}) which is consistent at the
1$\sigma$ level with the reverberation mapping-based mass of
$3.57^{+0.45}_{-0.37}\times 10^7~\msun$ (1$\sigma$ errors) obtained by
using data from \citet{bentz_etal_06}, but relying on a recently
updated empirical calibration of the RM mass scale by
\citet{grier_etal_13}. The best fit M/L ratio $\Upsilon_H = 0.34\pm
0.03$ (3$\sigma$ error) is consistent with the photometrically derived
mass-to-light ratio of $\Upsilon_H = 0.4{\pm 0.2}~\msun/\lsun$.
    
The new NIFS data yields $\sigma_c=116\pm3$\kms, the velocity
dispersion within $R_e/8$, which is $\sim$25\% larger than previous
values, while the new best estimate of $\mbh=3.76\pm1.15\times
10^7~\msun$ 17\% smaller than but fully consistent with both our
previous stellar dynamical upper limit \citep{onken_etal_07} and the
dynamical estimate based on H$_2$ line-emitting gas
\citep{hicks_malkan_08}. The larger value of $\sigma_c$ and smaller
value of $\mbh$ help to bring NGC~4151 closer to the best-fit \msigma
relation of \citet{woo_etal_13}.

The analysis in this paper demonstrates that biased estimates of BH
masses can arise when an axisymmetric orbit superposition code is used
to model a galaxy with a weak, but kinematically identifiable barred
galaxy, possibly resulting in an over-estimate in $\mbh$ if the M/L
ratio is constrained primarily by data beyond the sphere-of-influence
of the BH. This confirms the prediction made from the recent analysis
of $N$-body simulations of barred galaxies with BHs by
\citet{brown_etal_13}, and points to the need for new dynamical
modeling tools capable of modeling a stellar bar.  When such codes are
applied to the existing sample of barred galaxies, our results suggest
that it will likely enhance the discrepancy between barred and
unbarred galaxies in the $M_{\rm BH}-\sigma$ relation
\citep{graham_etal_11}.

Our modeling of this complex system suggests that the standard
practice of fitting kinematical constraints over a large range of
radii with a constant mass-to-light ratio \cite[for exceptions to
this practice, see][]{valluri_etal_05,mcconnell_etal_13} can bias the mass of the
central BH and the derived mass-to-light ratio. To some
extent, the biases can be overcome by only considering the kinematics
very close to the central BH, although this gives larger errors on the
estimated solutions of $\mbh$ and $\Upsilon_H$. Our tests of
robustness demonstrate that the use of such high spatial resolution
nuclear kinematical data in such modeling is very valuable. However,
the complications of the stellar dynamical bar imply that additional
work is required in order to perform the crucial test of the RM mass
scale calibration in this galaxy. Only with improved non-axisymmetric
modeling methods and/or BH mass measurements in other
reverberation-mapped AGNs will we ultimately be able to assess whether
the growth history of the BHs that we see accreting at present is
systematically different from those currently in quiescence.

\acknowledgments

We thank Remco van den Bosch for valuable input and for providing his
symmetrization routine. We thank Gaelle Dumas and Eric Emsellem for
providing us with the SAURON integral field kinematics. We thank Linda
Watson for providing copies of her NIFS stellar velocity template
spectra. We thank the anonymous referee for suggestions that helped to
improve our manuscript. We are grateful to Michele Cappellari for
making his analysis routines publicly available and for his continuing
efforts in adding software features. M.~Valluri and JB were supported
by National Science Foundation grant AST-0908346, and M.~Valluri
acknowledges support from the University of Michigan's Elizabeth
Crosby award. BMP and RWP are grateful to the National Science
Foundation for support of this work through grant AST-1008882 to The
Ohio State University. MB gratefully acknowledges support by the
National Science Foundation under grant AST-1253702.  M.~Vestergaard
acknowledges support from a FREJA Fellowship granted by the Dean of
the Faculty of Natural Sciences at the University of Copenhagen and a
Marie Curie International Incoming Fellowship.  The research leading
to these results has received funding from the People Programme (Marie
Curie Actions) of the European Union's Seventh Framework Programme
FP7/2007-2013/ under REA grant agreement No. 300553 (MV).  The Dark
Cosmology Centre is funded by the Danish National Research Foundation.

Based on observations obtained at the Gemini Observatory, which is
operated by the Association of Universities for Research in Astronomy,
Inc., under a cooperative agreement with the NSF on behalf of the
Gemini partnership: the National Science Foundation (United States),
the Science and Technology Facilities Council (United Kingdom), the
National Research Council (Canada), CONICYT (Chile), the Australian
Research Council (Australia), Minist\'{e}rio da Ci\^{e}ncia e
Tecnologia (Brazil) and Ministerio de Ciencia, Tecnolog\'{i}a e
Innovaci\'{o}n Productiva (Argentina). This research was supported in 
part through computational resources and services provided by Advanced 
Research Computing and the Flux cluster at the University of
Michigan, Ann Arbor. This work made use of data from
the Ohio State University Bright Spiral Galaxy Survey, which was
funded by grants AST-9217716 and AST-9617006 from the United States
National Science Foundation, with additional support from the Ohio
State University. Based on observations made with the NASA/ESA Hubble
Space Telescope, obtained from the data archive at the Space Telescope
Institute. STScI is operated by the association of Universities for
Research in Astronomy, Inc. under the NASA contract NAS 5-26555.
Funding for the SDSS and SDSS-II has been provided by the Alfred
P. Sloan Foundation, the Participating Institutions, the National
Science Foundation, the U.S. Department of Energy, the National
Aeronautics and Space Administration, the Japanese Monbukagakusho, the
Max Planck Society, and the Higher Education Funding Council for
England.  The SDSS is managed by the Astrophysical Research Consortium
for the Participating Institutions. The Participating Institutions are
the American Museum of Natural History, Astrophysical Institute
Potsdam, University of Basel, University of Cambridge, Case Western
Reserve University, University of Chicago, Drexel University,
Fermilab, the Institute for Advanced Study, the Japan Participation
Group, Johns Hopkins University, the Joint Institute for Nuclear
Astrophysics, the Kavli Institute for Particle Astrophysics and
Cosmology, the Korean Scientist Group, the Chinese Academy of Sciences
(LAMOST), Los Alamos National Laboratory, the Max-Planck-Institute for
Astronomy (MPIA), the Max-Planck-Institute for Astrophysics (MPA), New
Mexico State University, Ohio State University, University of
Pittsburgh, University of Portsmouth, Princeton University, the United
States Naval Observatory, and the University of Washington. This
research has made use of the NASA/IPAC Extragalactic Database (NED)
which is operated by the Jet Propulsion Laboratory, California
Institute of Technology, under contract with the National Aeronautics
and Space Administration. This research has made use of the SIMBAD
database, operated at CDS, Strasbourg, France.

{\it Facilities:} \facility{Gemini:Gillett (NIFS)}, \facility{MMT
  (Blue Channel spectrograph)}, \facility{Mayall}, \facility{Perkins},
\facility{HST (ACS/HRC)}, \facility{Sloan}

\end{document}